\documentclass[12pt]{article}
\usepackage{amsmath}
\usepackage{amssymb}
\usepackage{graphicx}
\usepackage{setspace}
\usepackage{graphicx}
\usepackage{epsfig}
\usepackage{amsfonts}
\usepackage{dsfont}
\usepackage{bm}
\usepackage{color}
\usepackage{comment}
\usepackage[dvipsnames]{xcolor}
\usepackage[authoryear,round]{natbib}
\usepackage{url}
\usepackage{pgfplots}
\usepackage[hidelinks]{hyperref}
\hypersetup{
  colorlinks   = true, %Colours links instead of ugly boxes
  urlcolor     = blue, %Colour for external hyperlinks
  linkcolor    = blue, %Colour of internal links
  citecolor   = blue %sepia %Colour of citations
}
\usepackage[T1]{fontenc}
\usepackage[cal=boondoxo]{mathalfa}

\usepackage{soul}

\newtheorem{innercustomthm}{Lemma}
\newenvironment{customthm}[1]
{\renewcommand\theinnercustomthm{#1}\innercustomthm}
{\endinnercustomthm}

\newtheorem{assumption}{\bf Assumption}
\newtheorem{proposition}{\bf Proposition}

\newenvironment{proof}[1][Proof]{\noindent\textbf{#1} }{\ \rule{0.5em}{0.5em}}

\pgfplotsset{compat=1.18}
\DeclareMathOperator{\dd}{\textrm{d}\!}
\allowdisplaybreaks

\begin{document}
\title{Tournaments with a Standard\thanks{We thank Alessandro Bonatti, Jennifer Brown, Andrew Rhodes, Sandro Shelegia and the participants of various seminars and conferences for helpful comments.}}
\author{Mikhail Drugov\thanks{Universitat Autònoma de Barcelona and Barcelona School of Economics (BSE), New Economic School and CEPR, \url{mdrugov@nes.ru}.} \and Dmitry Ryvkin\thanks{Economics Discipline Group, School of Economics, Finance and Marketing, RMIT University,  \url{d.ryvkin@gmail.com}.} \and Jun Zhang\thanks{Economics Discipline Group, School of Business, University of Technology Sydney, \url{jun.zhang-1@uts.edu.au}.}}
\date{This version: \today}

\maketitle

\begin{abstract}
\noindent We study tournaments where winning a rank-dependent prize requires passing a minimum performance standard. We show that, for any prize allocation, the optimal standard is always at a \emph{mode} of performance that is weakly higher than the global mode and identify a necessary and sufficient condition for it to be at the global mode.
%\textcolor{cyan}{we also have condition that the global mode is the largest, or the only one. ``necessary and sufficient condition'' sounds like very detailed for the abstract and it would be justified of our condition was very deep/illuminating but I am not sure it is the case} \textcolor{red}{I think it's sufficiently nontrivial that if the global mode is optimal for WTA then it's optimal for any prizes.} 
When the prize scheme can be designed as well, the winner-take-all prize scheme is optimal for noise distributions with an increasing failure rate; and awarding equal prizes to all qualifying agents is optimal for noise distributions with a decreasing failure rate. For distributions with monotone likelihood ratios---log-concave and log-convex, respectively---these pay schemes are also optimal in a larger class of anonymous, monotone contracts that may depend on cardinal performance.

\bigskip


\end{abstract}

\newpage
\onehalfspacing

\section{Introduction}

In many tournaments, meeting a minimum performance standard is necessary to qualify for a prize. For example, in US universities, the allocation of salary raises or bonuses is often based on performance rankings. Typically, an executive, such as a department chair or a dean, has a fixed pool of money to allocate, but only faculty members who at least ``meet expectations'' are eligible to receive a raise.\footnote{For example, the HR guidelines at UCLA state (\url{https://chr.ucla.edu/training-and-development/performance-management-tools-resources}): ``Staff members with a performance rating of `meets expectations' or above as determined in the most recent performance appraisal are eligible for merit increase consideration.''} How should the standard be optimally set? How does it operate in conjunction with the prize scheme used in the tournament?

In this paper, we study how a minimum performance standard, which we call simply \textit{standard}, should be set by a tournament organizer who cares about the agents' effort. While such standards are common in many environments with tournament incentives,\footnote{In innovation tournaments, standards can take the form of a fixed challenge (an unsolved problem), or a requirement that the innovation satisfies certain specifications or sufficiently improves upon an existing product. In 1996, the XPRIZE Foundation offered the Ansari X Prize for the first non-government organization to launch a reusable crewed spacecraft into space twice within two weeks (\url{https://space.xprize.org/prizes/ansari}). The 2021 XPRIZE Carbon Removal initiative (\url{https://www.xprize.org/prizes/carbonremoval}), with a \$100M prize purse to be distributed between one top prize of \$50M and three secondary prizes of \$10M each (plus additional smaller prizes for student teams), has a minimum standard of 1000 tonnes of CO2 removed per year for working solution. The Netflix Prize (\url{https://www.netflixprize.com/rules.html}) was set up in 2006 for the best algorithm predicting user ratings of films, under the condition that it had to be at least 10\% more accurate than Netflix's existing algorithm at the time. The 2004 DARPA Grand Challenge was a competition for the fastest driverless car to complete a 142-mile course \citep{Hooper:2004}.} their effect has not been systematically studied. We use a rank-order tournament model \`{a} la \cite{Lazear-Rosen:1981} where players' performance is their effort distorted by idiosyncratic additive noise, and focus on risk-neutral, homogeneous players. We find a surprising degree of universality in the choice of the optimal standard: For \textit{any} allocation of prizes, the effort-maximizing standard is at one of the \emph{modes} of players' performance.\footnote{Consequently, for a unimodal density of performance a standard at the mode is always optimal.} That is, the optimal standard mimics a modal competitor. 

For intuition, suppose first that the density of noise is unimodal and consider a \emph{winner-take-all} (WTA) tournament where the entire prize budget is awarded to the top performer. In a symmetric equilibrium, a player's incentives at the margin are determined by the likelihood that his noise realization coincides with the maximum between (i) the largest realization of noise among all other players and (ii) the \textit{threshold level of noise}, defined as the standard performance less the equilibrium effort. Therefore, setting the threshold level of noise at the mode of the noise distribution or, equivalently, setting the standard at the mode of performance, maximizes marginal incentives.

Interestingly, a similar intuition applies to more complex prize schedules where weakly declining prizes are awarded to several top performers. Such prize schedules are common in many environments, such as salary raises in Universities as mentioned above as well as sales or innovation contests.\footnote{For example, ``21 Sales Competition Ideas That Will Fire Up Your Team'' on \url{ambition.com}, a sales coaching platform, lists as idea \#3 ``Don't always lean on `first place winner' contests. Move outside the box to engage more of your middle.'' Innovation contests, such as those run by the XPRIZE Foundation, coding contests run by TopCoder, procurement contests run by the US federal government, as well as sports competitions quite often involve multiple prizes. For further examples, see, e.g., \cite{Halac-et-al:2017,Fang-et-al:2020,Drugov-Ryvkin:2020}.} There, similar to the WTA case, incentives created by a positive \emph{prize differential} at some rank depend on the likelihood that the player's noise realization coincides with the noise realization of a competitor for that rank or the threshold level of noise, whichever of the two is higher. Therefore, marginal incentives \emph{for each rank} are maximized when the threshold is set at the mode of noise. One implication is that if the distribution of noise is positively skewed to the extent that its mode is at the lower bound of the support, all agents will pass the optimal standard in equilibrium. For example, almost all university faculty typically ``meet expectations'' in their performance reviews and are eligible for raises. When the mode is not at the lower bound it is, of course, possible that none of the agents reaches the standard performance, in which case nobody receives any prize.\footnote{For example, the best-performing vehicle of Carnegie Mellon University's \textit{Red Team} completed only 7.32 miles out of 142 in the 2004 DARPA Grand Challenge. Notably, DARPA described the purpose of its challenge as ``to leverage American ingenuity to accelerate the development of autonomous vehicle technologies that can be applied to military requirements'' (\url{http://archive.darpa.mil/grandchallenge04/overview.htm}). The goal, therefore, was not to solve a specific problem but to boost innovation effort in a particular sector.} 

The intuition that the optimal standard mimics a modal competitor extends also to the mutlimodal case.\footnote{Multimodal distributions typically arise due to mixing of several unimodal populations. They have been identified empirically in many contexts, from preferences and income of consumers \citep{Dixit-Weibull:2007,Fiorina-Abrams:2008,Autor-et-al:2006} to macroeconomic variables \citep{Brunnermeier-Sannikov:2014}, medical costs \citep{Patterson:2011}, mortality times \citep{Gunst-et-al:2010}, failure times of devices \citep{Fischer-et-al:2000}, and regulatory approval times for new drugs \citep{Grabowski-Vernon:1990}.} Here, it is optimal to set the standard at \emph{some} mode of performance, although which mode it is, in general, depends on the details of the distribution of noise and on the prize schedule. The key observation is that more unequal prize schedules are associated with stricter standards (i.e., standards that are harder to pass in equilibrium). %in terms of threshold levels of noise. \textcolor{green}{Do we have the above result?} \textcolor{cyan}{yes, see fn. 20} \textcolor{green}{Also, stricter standards sound like higher standards, which is not true in general, only the noise threshold is higher.} \textcolor{cyan}{I added above ``in terms of threshold levels of noise'' - is this OK?} \textcolor{red}{There's a reason it says "stricter" and not "higher" and also "associated" and not "always resu;t from" or similar. These are intuitive connections that don't need to be precise at this point. }
In particular, the WTA tournament, which is the most unequal, calls for a standard with the largest threshold level of noise. At the opposite end of the spectrum is the \emph{equal prize sharing} (EPS) tournament in which all qualifying agents receive a fixed payment, and the optimal threshold is the lowest. In the latter case there is no competition, i.e., players' incentives are determined only by the standard, and hence the optimal threshold is at the \emph{global mode}---the global maximum of the density of noise. Thus, for any prize schedule it is optimal to set the standard at a mode of performance that is (weakly) above the global mode. Moreover, if a standard at the global mode is optimal in the WTA case, it is also optimal for any prize allocation.

Having established the properties of optimal standards for arbitrary prize schedules, we also study how to optimally allocate prizes, restricting attention to settings where the standard at the global mode is optimal. The answer depends critically on the properties of noise. When the distribution of noise has an increasing failure (or hazard) rate (IFR) after the global mode, the WTA schedule is optimal. Many popular ``light-tailed'' distributions such as normal, uniform, or logistic are IFR---and we require a weaker condition to be IFR after the global mode. In contrast, for decreasing failure rate (DFR) distributions of noise, such as the Pareto distribution, which exhibit ``heavy tails'', it is optimal to share the prize equally among all the players who pass the standard. Moreover, since the density of noise is decreasing in this case, the mode is at the lower bound of the support, and hence exerting the equilibrium effort \textit{guarantees} passing the standard in equilibrium. Thus, under the optimal pay scheme in the DFR case, there is effectively no competition among the players, and the tournament can be replaced by individual bonuses which are paid to all who qualify. If the prize budget is indivisible, this prize scheme can nevertheless be implemented by randomly awarding one prize---the entire budget---to one of the players who passed the standard. 

It has long been argued \citep[see, e.g., the discussion in the original article by][]{Lazear-Rosen:1981} that an important advantage of tournament pay schemes is that they only rely on \textit{ordinal} performance comparisons, as opposed to standard moral hazard contracts that are based on cardinal information.\footnote{It may be argued that in the presence of a cardinal standard a tournament pay scheme ceases to be fully ordinal. However, it is still ordinal in the sense that the only additional assessment the principal needs to make is whether or not output \textit{passes} the standard---a qualitative assessment, which is arguably not more demanding than deciding whether one player's output is higher than another player's output.} Therefore, tournaments can be used in settings where cardinal performance measurements are unreliable and cannot be contracted on, but, generally speaking, the principal should be able to do better when cardinal information is available. We show, however, that when noise is either log-concave or log-convex---which are slightly stronger conditions than IFR and DFR, respectively---the principal \textit{cannot} do better than using the corresponding optimal tournament with a standard. In other words, the optimal tournament-with-a-standard contracts we identify are optimal in a much larger class of pay schemes. Note that the optimal moral hazard contract even for a single agent is unknown when the monotone likelihood ratio property (MLRP) does not hold, i.e., in our setting, the distribution of noise is not log-concave. Our results provide a solution for the case of multiple agents and log-convex noise.

Finally, we apply our results to other prominent models of noisy contests---the \cite{Tullock:1980} contest, the innovation contest of \cite{Fullerton-McAfee:1999}, and the patent race of \cite{Loury:1979}. The underlying microfoundations of these models allow us to explore the effect of a standard and identify the optimal standard and optimal pay scheme in each case.

\paragraph{Related literature}
Starting with \cite{Lazear-Rosen:1981}, a number of authors investigated the problem of optimal prize allocation in rank-order tournaments. Yet, while already  \cite{Lazear-Rosen:1981} mentioned competing against a ``fixed standard'' as a way to obtain the first-best effort level, this literature has so far ignored the analysis of optimal standards of performance and associated prize schemes.\footnote{In the logit random utility model, which is formally related to the Tullock contest, the outside option---effectively, a standard---has been introduced a long time ago \citep{Anderson-et-al:1992}. Yet, its optimality has not been studied because it is treated as a parameter. Also, log-concavity \citep[or a slightly weaker condition of $-1/(n+1)$-concavity, see][]{Caplin-Nalebuff:1991} has to be assumed for technical reasons in that model.} For symmetric, unimodal noise distributions, \cite{Krishna-Morgan:1998} show the optimality of WTA in small tournaments (up to four risk-neutral players or up to three risk-averse players). \cite{Kalra-Shi:2001} and \cite{Terwiesch-Xu:2008} show that WTA is optimal for some particular log-concave distributions of noise. \cite{Schweinzer-Segev:2012} show the optimality of WTA for Tullock contests.\footnote{The multi-prize nested Tullock contest can be equivalently represented as a rank-order tournament with the Gumbel distribution of noise, which is log-concave \citep{Jia:2008,Fu-Lu:2012,Ryvkin-Drugov:2020}.} \cite{Akerlof-Holden:2012} and \cite{Ales-et-al:2017} make somewhat different modeling assumptions and do not state most results in terms of the primitives; the latter paper, however, shows the crucial role of log-concavity of the noise distribution for the optimality of the WTA prize scheme. \cite{Drugov-Ryvkin:2020} and \cite{Drugov-Ryvkin:2022} extend and generalize many of the previous results for the risk-neutral and risk-averse case, respectively.

The minimum performance standard in our setting resembles a reserve price in auctions, which has been studied extensively starting with \cite{Myerson:1981} and \cite{Riley-Samuelson:1981}, including all-pay auctions in the context of innovation \citep{Erkal-Xiao:2021}. However, the role of a reserve price in auctions is different from that of a standard in tournaments. For example, in an independent private value auction setting, the optimal reserve price excludes low types as long as the effect of the FOSD shift in the distribution of types outweighs the reduction in competition due to a decrease in the number of bidders. In a (symmetric) tournament, we can think of noise realizations as ``types,'' except they are determined after bidding, i.e., a performance standard excludes low types \textit{ex post}. Therefore, there is no reduction in competition due to a decrease in the number of bidders, but rather there is some effect on marginal incentives because a player may not qualify for a prize even when he is highly ranked if he cannot pass the standard.\footnote{\cite{Bulow-Klemperer:1996} show that in an auction adding another player is always better than introducing a reserve price. In our setting, this comparison is ambiguous and depends on the noise distribution. Hence, the standard in tournaments might benefit the principal more than the reserve price in auctions.}
 
Finally, our paper is related to the literature on contracting under moral hazard. Since the classical papers of \cite{Holmstrom:1979}, \cite{Rogerson:1985} and \cite{Jewitt:1988}, it was understood that the shape of optimal contracts follows that of the likelihood ratio---the log-derivative of the density of output with respect to effort. In particular, only positive likelihood ratios are to be rewarded under limited liability, which gives rise to a minimum performance standard. However, the monotone likelihood ratio property (MLRP)---which is equivalent to log-concavity of the noise distribution in our setting---has been routinely imposed in this literature to ensure that the monotonicity constraint for optimal contracts is not binding in the principal's problem. %Other technical conditions, such as the convexity of the distribution function condition (CDFC), have been imposed for the sufficiency of the first-order approach in the agent's problem. 
There has been some progress recently in analyzing settings where MLRP does not hold \citep[e.g.][]{Poblete-Spulber:2012,Kadan-Swinkels:2013,Kirkegaard_TE:2017}. Still, optimal contracts when (additive) noise is not log-concave have not been discussed---even in the single agent case. A recent study by \cite{Kirkegaard:2023} connects the contest literature to the moral hazard literature by considering contests with heterogeneous players in a general moral hazard setting. Under the MLRP, %and CDFC, 
\cite{Kirkegaard:2023} shows that optimal contest design includes a minimum performance standard that corresponds, as noted above, to the zero of the likelihood ratio. Our model is more specialized in some aspects and more general in others: We restrict attention to rank-order tournaments with additive noise, but allow for arbitrary prize allocations and noise distributions that do not have a monotone or even single-crossing likelihood ratio. %We also do not impose the CDFC but use the strict convexity of the cost function to help with the first-order approach. 
We show that the resulting characterization of optimal contracts does not rely on the shape of the likelihood ratio at all, even though the results of \cite{Kirkegaard:2023} are recovered when the MLRP is imposed. We discuss this connection in detail in Section \ref{sec:opt_cardinal}.

\bigskip

The rest of the paper is organized as follows. Section \ref{sec:model} sets up the model. In Section \ref{sec:tournaments}, we present our mains results: first, on the optimality of a standard at the modal performance for arbitrary prize schedules, and second, the optimal prize schedules. In Section \ref{sec:opt_cardinal}, we show that in many cases tournaments with a standard are optimal in a class of pay schemes based on cardinal output. In Section \ref{sec:applications}, we discuss the connection between our setting and three prominent contest models and the implications of our results for those models. Section \ref{sec:conclusion} concludes. All missing proofs are collected in Appendix \ref{app_proofs}.

\section{The Model}
\label{sec:model}

There are $n\ge 2$ symmetric, risk-neutral players, indexed by $i=1,\dots,n$, who simultaneously and independently choose effort levels $e_i\in\mathds{R}_+$. Player $i$'s performance is stochastic and given by $Y_i=e_i+X_i$, where shocks $X_i$ are i.i.d. across players, with distribution $F(\cdot)$ and density $f(\cdot)$ supported in an interval $\mathcal{X}=[\underline{x},\overline{x}]$ (where $\underline{x}<\overline{x}$ can be finite or infinite). We assume that $f$ is continuous, bounded, piece-wise differentiable in $\mathcal{X}$, and satisfies $f(\overline{x})=0$.\footnote{The last condition holds automatically when $\overline{x}=+\infty$. For a finite $\overline{x}$, it is adopted mainly for the ease of exposition and is not required for many of our results. It is critical, however, for the log-convex case where it ensures that $f$ is decreasing and has a decreasing failure rate (DFR). Note that we do \textit{not} assume $f(\underline{x})=0$, which would exclude many important cases, e.g., DFR distributions.} Let $x_m$ denote the largest global maximizer of $f(\cdot)$. We allow for $f$ to be \emph{multimodal} and refer to $x_m$ as the \emph{global mode} of $f$. The cost of effort $e_i$ to player $i$ is $c(e_i)$, where $c(\cdot)$ is strictly increasing from zero, and $\bar{e}=c^{-1}(1)$---the largest undominated effort in a tournament with unit prize budget---is finite. Furthermore, $c(\cdot)$ is $C^1$ and strictly convex on $[0,\bar{e}]$, $C^2$ on $(0,\bar{e}]$, and satisfies $c(0)=c'(0)=0$.

A (tournament-with-a-standard) pay scheme, $(\rho,{\bf v})$, consists of a standard $\rho\in \mathds{R}$ and a prize schedule ${\bf v}\in \mathcal{V}=\{{\bf v}\in\mathds{R}^n_+:v_1\ge\ldots\ge v_n\ge 0,\,\sum_{r=1}^nv_r=1\}$, where prizes are decreasing, non-negative, and (without loss) the total prize budget is normalized to one. The principal observes whether or not each player's performance passes the standard,\footnote{We use terms such as ``pass'' or ``above'' in the weak sense. Meeting the standard is considered passing, but happens with probability zero for an atomless $F$.} as well as the ranking of performance above the standard. If none of the players passes the standard, no prizes are awarded; otherwise, prize $v_r$ is awarded to the player with performance ranked $r$ among the players above the standard. That is, the player with the highest performance gets $v_1$, the second highest gets $v_2$, etc. There is no need to specify a tie-breaking rule as ties occur with probability zero for an atomless $F$.

Two simple prize schedules in $\mathcal{V}$ deserve special consideration: (i) the \textit{winner-take-all} (WTA) prize schedule, ${\bf v}^{\rm WTA}=(1,0,\ldots,0)$, whereby the entire prize budget is awarded to the top performer (if he passed the standard); and (ii) the \textit{equal prize sharing} (EPS) schedule,  ${\bf v}^{\rm EPS}=(\frac{1}{n},\ldots,\frac{1}{n})$, which awards $\frac{1}{n}$ to all the players who pass the standard. These prize schedules represent the two extremes in terms of inequality of outcomes, with the WTA schedule being the most unequal, and the EPS schedule the most equal in $\mathcal{V}$.\footnote{These notions can be formalized by (partially) ranking prize schedules in the \textit{majorization order} \citep{Marshall-et-al:2011}. The WTA schedule majorizes all others in $\mathcal{V}$, while the EPS schedule is majorized by all others in $\mathcal{V}$.} They can also be viewed as extreme in terms of the level of difficulty or competitive pressure: In order to win a prize under WTA a player has to be the top performer and pass the standard, whereas to win a prize under EPS it is sufficient to only pass the standard and, effectively, there is no competition among the players.

The principal's objective is to maximize total effort, $\sum_{i=1}^ne_i$, which in our setting is equivalent to maximizing total expected performance, $\mathds{E}(\sum_{i=1}^nY_i)$. Furthermore, we restrict attention to the implementation of a symmetric, pure strategy equilibrium where all players $i=1,\ldots,n$ choose the same effort $e_i=e^*$. The principal's objective is then also equivalent to maximizing individual effort, $e^*$.\footnote{Thus, prize payments are not part of the principal's objective. Practically speaking, we assume that the principal values prizes little, or not at all, relative to effort or output. This holds in our leading example of rank-based raises or bonuses where administrators are allocated a fixed budget for such raises that cannot be spent on other things, and any residual funds are swept away at the end of the budgeting period. This occurs also in educational settings where instructors value students' effort regardless of the grades they eventually assign, and those grades are costless. Competition for status (e.g., rankings or medals in sports) has similar features. Likewise, a research funding agency may be interested in boosting research output and innovation regardless of whether grants are eventually awarded or prizes are paid out. If this assumption is relaxed and prize payments are part of the principal's objective, we can show that the optimal standard is higher than the one we identify in Section \ref{sec:opt_standard}.} 

\section{Optimal Tournaments with a Standard}
\label{sec:tournaments}

\subsection{The Optimal Standard}
\label{sec:opt_standard}

In this section, we assume prize schedule ${\bf v}\in\mathcal{V}$ is fixed and characterize the optimal standard. Focusing on a symmetric pure strategy equilibrium, suppose all players except one choose some effort $e^*$. Then, for a given standard $\rho$, we can write the expected payoff of the deviating player with effort $e$ as
\begin{align}
\label{payoff_general}
\pi(e,e^*;\rho) = \sum_{r=1}^np^{(r)}(e,e^*;\rho)v_r - c(e),
\end{align}
where $p^{(r)}(e,e^*;\rho)$ is the probability for the deviating player to receive prize $v_r$. Using summation by parts, we rewrite (\ref{payoff_general}) as
\begin{align}
\label{payoff_byparts}
\pi(e,e^*;\rho) = \sum_{r=1}^nP^{(r)}(e,e^*;\rho)(v_r-v_{r+1}) - c(e),
\end{align}
where $P^{(r)}(e,e^*;\rho) = \sum_{k=1}^rp^{(k)}(e,e^*;\rho)$ is the probability of receiving a prize of \emph{at least} $v_r$. Thus, the deviating player's payoff is represented as a linear combination of non-negative \emph{prize differentials} $v_r-v_{r+1}$ (where, slightly abusing notation, we define $v_{n+1}=0$). Each prize differential is the gain for the player from being able to secure a prize of $v_r$ or better as compared to $v_{r+1}$ or better, and hence it is multiplied by $P^{(r)}(e,e^*;\rho)$.

Probability $P^{(r)}(e,e^*;\rho)$ can be written as
\begin{align}
\label{P_r}
& P^{(r)}(e,e^*;\rho) = \mathds{P}(e+X>\max\{\rho,e^*+X_{(n-r:n-1)}\}) \nonumber\\
& = \bar{F}(\rho-e)F_{(n-r:n-1)}(\rho-e^*) + \int\limits_{\{x>\rho-e^*\}} \bar{F}(e^*-e+x)\dd F_{(n-r:n-1)}(x).
\end{align}
Here, $X$ is the deviating player's noise realization, $X_{(n-r:n-1)}$ is the noise realization ranked $r$ among $n-1$ i.i.d. draws from $X$,\footnote{We follow the standard enumeration of order statistics \citep[e.g.,][]{David-Nagaraja:2004} and use subscript $(j:n)$ to denote random variables and distributions corresponding to the $(n+1-j)$-th highest draw in a sample of size $n$. To simplify notation, we also adopt the convention that $X_{(0:n)}=-\infty$ and $F_{(0:n)}(x)=1$ for all $x\in\mathds{R}$.} and $\bar{F}(\cdot)=1-F(\cdot)$ is the survival function of noise. Indeed, the deviating player earns at least $v_r$, for $r\le n-1$, if he passes the standard \emph{and} his performance exceeds that of the player ranked $r$ among the remaining $n-1$ players; and the player earns at least $v_n$---the last prize---if he simply passes the standard.

Representation (\ref{P_r}) highlights two incentive effects in the tournament with a standard. The first term corresponds to scenarios where $e^*+X_{(n-r:n-1)}<\rho$, i.e., the realization of $X_{(n-r:n-1)}$ is too low for the competitor ranked $r$ to pass the standard. In this case, the deviating player's incentives for prize differential $v_r-v_{r+1}$ are determined entirely by his own probability of passing the standard, $\bar{F}(\rho-e)$. The second term covers scenarios where $X_{(n-r:n-1)}$ is large enough so that the competitor ranked $r$ passes the standard, and hence in order to gain $v_r-v_{r+1}$ the deviating player needs to surpass this competitor, which happens with probability $\bar{F}(e^*-e+x)$, for a given $X_{(n-r:n-1)}=x$. The former effect vanishes for $\rho\to-\infty$ (when there is no standard) and for $\rho\to +\infty$ (when the standard is so high that it can never be passed); therefore, there is an interior standard $\rho$ where this effect is maximized. The latter effect disappears for $r=n$ (i.e., when there is no competitor to outperform) and is decreasing monotonically with $\rho$.

We adopt the first-order approach, i.e., assume that the symmetric first-order condition (FOC), $\pi_e(e^*,e^*;\rho)=0$, holds.\footnote{\label{fn eq existence}To ensure the equilibrium existence and uniqueness we require that (i) function $\pi(e,e^*;\rho)$ is strictly concave in $e$ for $e\in[0,\bar{e}]$; (ii) the symmetric FOC $\pi_e(e^*,e^*;\rho)=0$ has a unique solution $e^*>0$; (iii) the participation constraint $\pi(e^*,e^*;\rho)\ge 0$ is not binding. Sufficient conditions for (i)-(iii) in terms of the primitives can be formulated assuming $\inf_{e\in[0,\bar{e}]}c''(e)>0$ and the distribution of noise is sufficiently dispersed. For a similar approach, see \cite{Drugov-Ryvkin:2020}. We eschew the details here for the ease of exposition.} The marginal benefit of effort in the symmetric equilibrium is then determined by coefficients $B_r(\rho-e^*)=P^{(r)}_e(e^*,e^*;\rho)$, where, for any $t\in \mathds{R}$, $B_r(t)$ is obtained by differentiating (\ref{P_r}) with respect to $e$ and setting $e=e^*$:
\begin{align}
\label{B_r}
B_r(t) = f(t)F_{(n-r:n-1)}(t) + \int_{\{x>t\}}f(x)\dd F_{(n-r:n-1)}(x).
\end{align}
The symmetric FOC, therefore, takes the form
\begin{align}
\label{FOC_main}
g(\rho-e^*;\mathbf{v}) = c'(e^*),
\end{align}
where 
\begin{align}
\label{g_general}
g(t;{\bf v}) = \sum_{r=1}^nB_r(t)(v_r-v_{r+1})
\end{align}
is the total marginal benefit of effort when the standard is $\rho=e^*+t$.

Recall that $c(\cdot)$ is strictly convex. It follows directly from (\ref{FOC_main}) that the highest $e^*$ that can be achieved corresponds to the global maximum of $g(t;\mathbf{v})$ with respect to $t$, if one exists. This leads to our first result about a general characterization of the optimal standard.

\begin{proposition}
\label{prop_opt_reserve}
For any prize schedule $\mathbf{v}\in\mathcal{V}$, it is optimal to set the standard at $\rho^*=e^*+t^*$, where $t^*$ is a mode of $f$ that is (weakly) above the global mode, and the equilibrium effort is $e^*=c'^{-1}(g(t^*;\mathbf{v}))$.
\end{proposition}

Proposition \ref{prop_opt_reserve} exploits the separable structure of performance in the model with additive noise.\footnote{With proper modifications, a similar result can be obtained for multiplicative noise.} The optimal standard is characterized in two steps. First, the optimal threshold level of noise, $t^*$, is determined by the distribution of noise and the prize schedule. Second, the resulting equilibrium effort, $e^*$, is such that its marginal benefit, $g(t^*;{\bf v})$, is equal to the marginal cost, and the optimal standard is $e^*+t^*$. The first step is independent of the equilibrium effort and allows us to interpret the standard in terms of the corresponding realizations of noise. For example, $\bar{F}(t^*)$---the probability of passing the optimal standard, and hence how \emph{strict} the optimal standard is---can be identified without knowing $e^*$. 

\begin{figure}
\centering
{%
\includegraphics[width=3in]{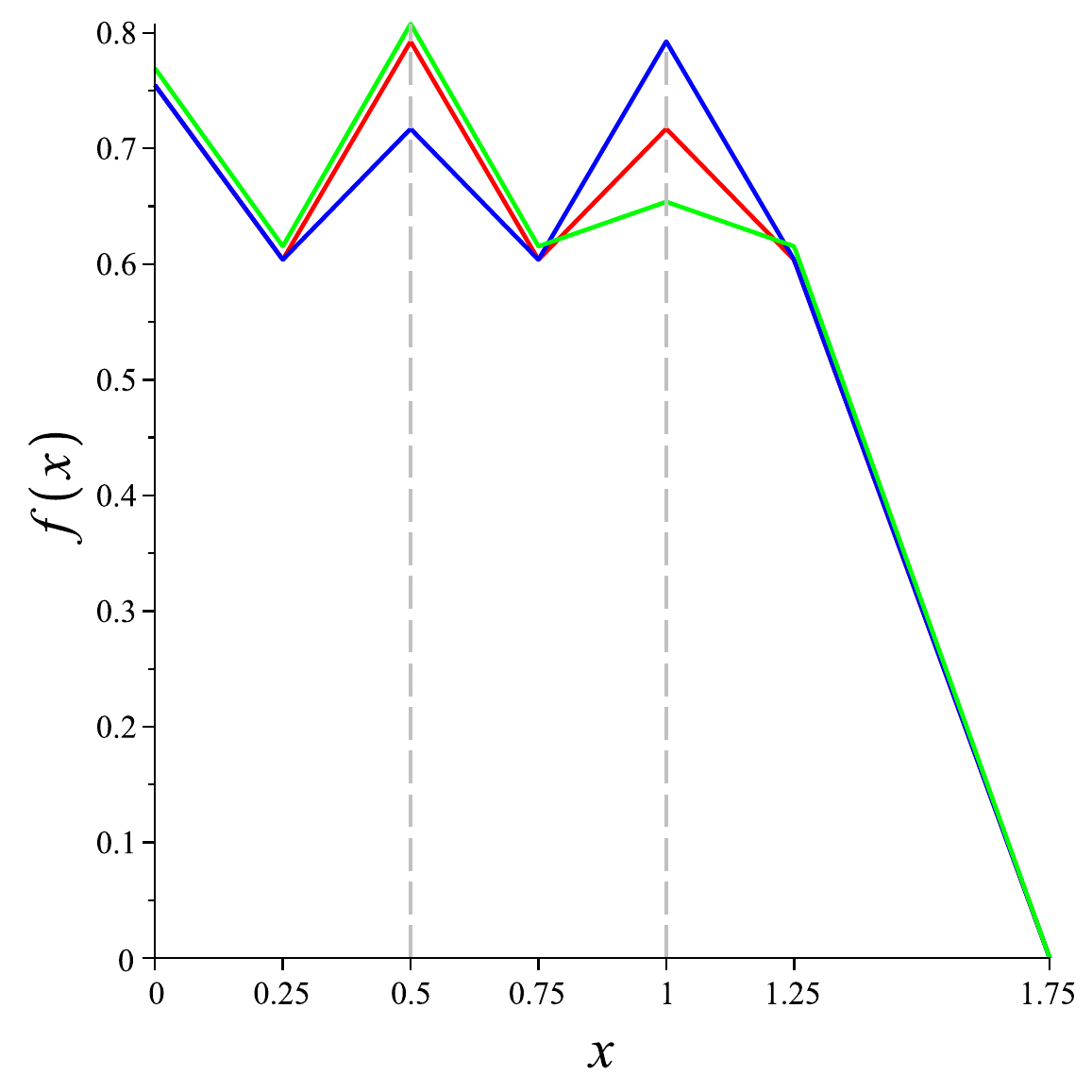}
}
\hfill
{%
\includegraphics[width=3in]{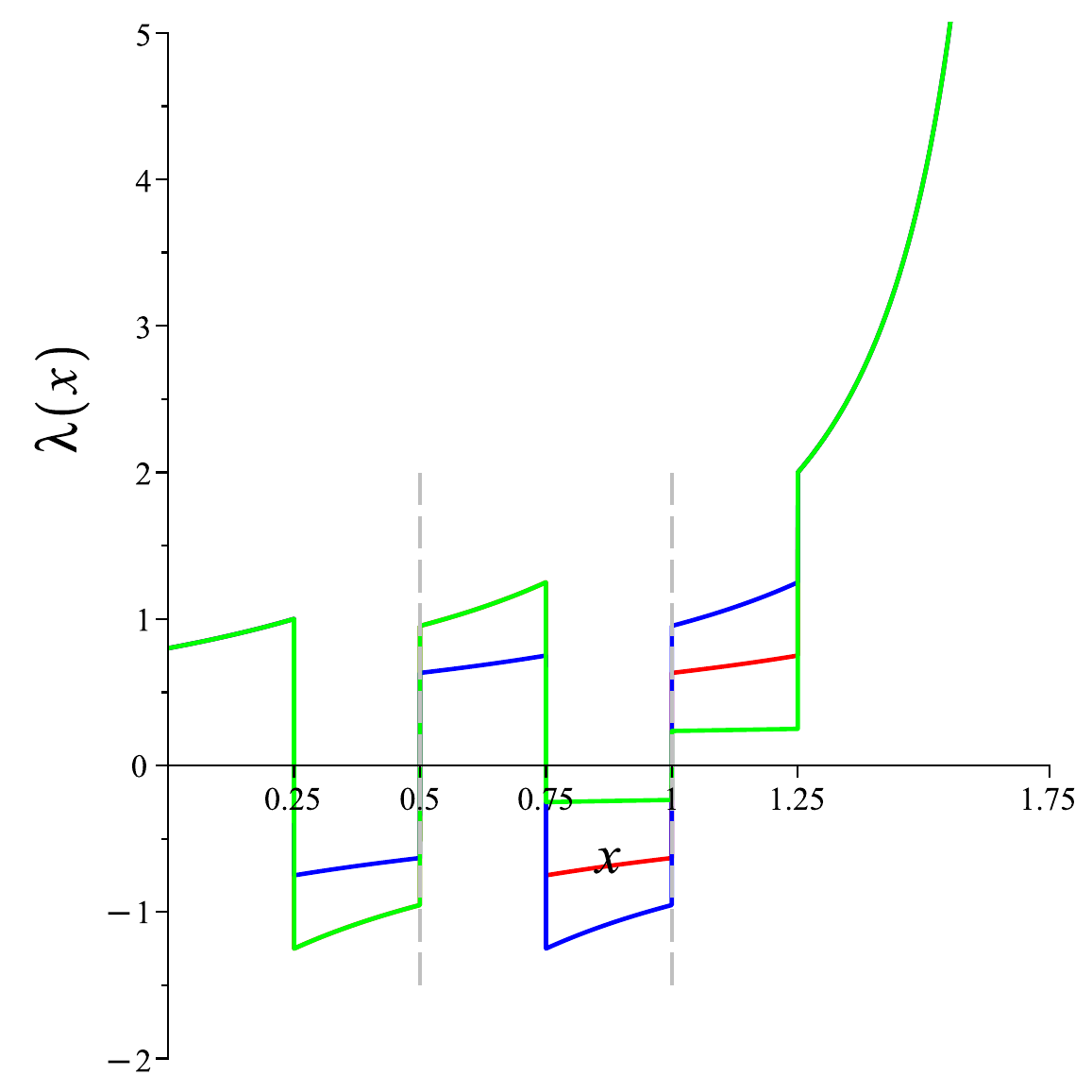}
}
{%
\includegraphics[width=3in]{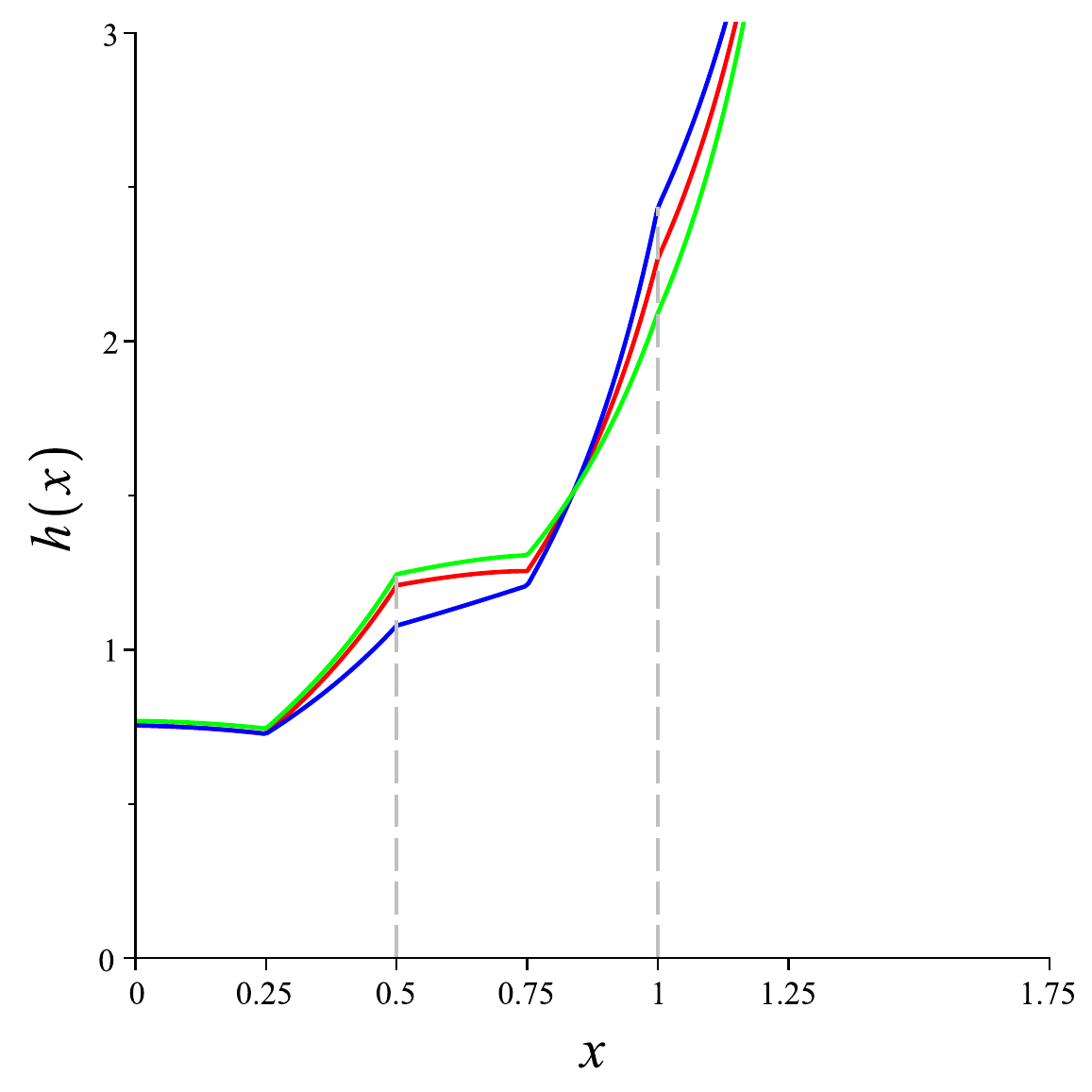}
}
\hspace{10pt}
{%
\includegraphics[width=3in]{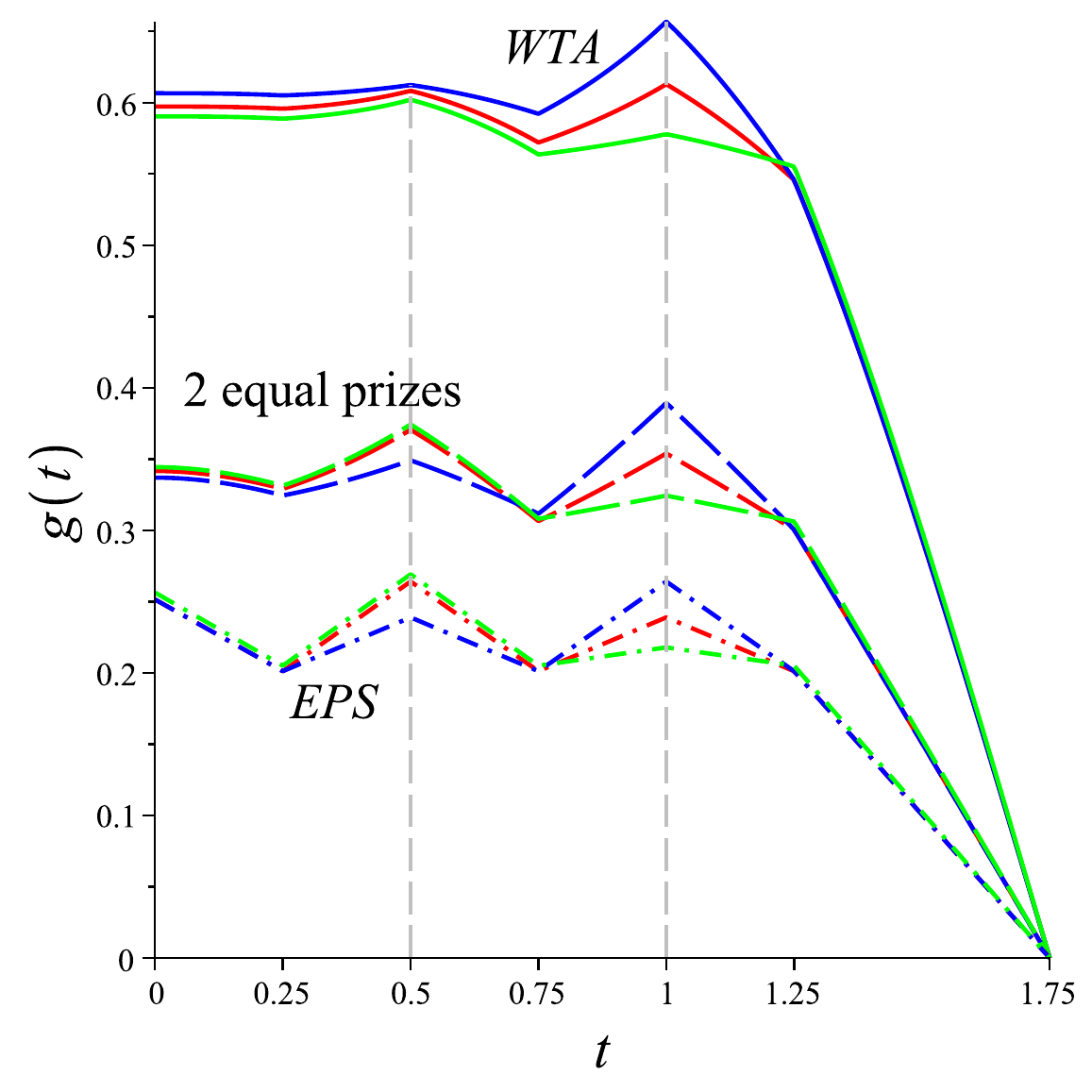}
}
\caption{\small{The density $f$ (top left), likelihood ratio $\lambda=-f'/f$ (top right), hazard rate $h=f/(1-F)$ (bottom left) and function $g(t;{\bf v})$ (bottom right, for different prize schedules ${\bf v}$) for three distributions with piecewise linear densities on $[0,1.75]$. The distribution shown in red has \\ $f(x) =  \frac{1}{16} \left\{ \begin{array}{ll}
20-16x & \text{if } 0 \leq x < 0.25 \\
11+20x & \text{if } 0.25 \leq x < 0.5 \\
31-20x & \text{if } 0.5 \leq x < 0.75 \\
7+12x & \text{if } 0.75 \leq x < 1 \\
31-12x & \text{if } 1 \leq x < 1.25 \\
56-32x & \text{if } 1.25 \leq x \leq 1.75
\end{array} \right.$
\\  and the two others are constructed similarly. Parameters: $n=3$; ${\bf v}=(1,0,0)$ for WTA, $(1/2,1/2,0)$ for two equal prizes, and $(1/3,1/3,1/3)$ for EPS.}}
\label{fig:example_main}
\end{figure}

Proposition \ref{prop_opt_reserve} implies immediately that the optimal standard is at the global mode of performance when it is also the largest mode. This trivially includes the important case of unimodal distributions. For a more general multimodal case, the proposition is illustrated in Figure \ref{fig:example_main} where the top left panel shows three trimodal distributions.\footnote{These distributions are inspired by an example from \cite{Barlow-et-al:1963} of a distribution that is IFR but not log-concave.} For each of the distributions, the bottom right panel shows function $g(t;\mathbf{v})$ for three different prize schemes. As seen in the figure, in all cases, $g(t;\mathbf{v})$ is maximized either at the global mode or at a higher mode.

The proof of Proposition \ref{prop_opt_reserve}, provided in Appendix \ref{app_proofs}, proceeds by rewriting coefficients $B_r(t)$ in (\ref{B_r}) in the form
\begin{align}
\label{B_r_f_tilde}
B_r(t) = \int \tilde{f}(x;t)\dd F_{(n-r:n-1)}(x), 
\end{align}
where $\tilde{f}(x;t) = f(\max\{x,t\})$. To interpret representation (\ref{B_r_f_tilde}), recall that $B_r(t)$ is the marginal effect of effort, in equilibrium, on the probability of a player's performance being ranked $r$ or higher and passing the standard $\rho=e^*+t$. Intuitively, a marginal increase in effort is pivotal for prize differential $v_r-v_{r+1}$ when the player's performance is close to the performance of the competitor ranked $r$ or the standard, whichever is higher. In the symmetric equilibrium, this happens when the player's noise realization, $X$, is close to $\max\{X_{(n-r:n-1)},t\}$. Therefore, marginal incentives are determined by the probability density of $X-\max\{X_{(n-r:n-1)},t\}$ at zero, which, using the convolution formula,\footnote{For independent random variables $X$ and $Y$ with densities $f_X$ and $f_Y$, respectively, the density of the difference $Z=X-Y$ is $f_Z(z) = \int f_X(y+z)f_Y(y)\dd y$, and hence $f_Z(0) = \int f_X(y)f_Y(y)\dd y = \mathds{E}(f_X(Y))$.} can be written as $B_r(t)=\mathds{E}(f(\max\{X_{(n-r:n-1)},t\})) = \mathds{E}(\tilde{f}(X_{(n-r:n-1)};t))$, producing (\ref{B_r_f_tilde}). For $r=n$, i.e., to at least receive the last prize, the player only needs to pass the standard and hence $B_n(t)=f(t)$, which also follows formally from (\ref{B_r}) or (\ref{B_r_f_tilde}).

While Proposition \ref{prop_opt_reserve} restricts the optimal standard to the modes at or to the right of the global mode of performance, the exact location of the standard remains ambiguous when the distribution of noise is multimodal. In particular, which mode is optimal depends on the prize schedule. For example, for the distribution shown in red in Figure \ref{fig:example_main}, the right mode $x=1$ is optimal for WTA but the left (and global) mode $x_m=0.5$ is optimal when there are two or three (EPS) equal prizes.

Representation (\ref{B_r_f_tilde}) shows that marginal benefits $B_r(t)$ for different ranks $r=1,\ldots,n$ can be maximized at different modes and thus, as seen from (\ref{g_general}), depending on the structure of prize differentials $v_r-v_{r+1}$ different modes can be ``activated'' to generate the total marginal benefit $g(t;{\bf v})$. One can think about a multimodal distribution of noise as a mixture of several ``subpopulations,'' each with its own mode, where the noise is drawn from \citep[this interpretation goes back at least to][]{Pearson:1894}. These correspond to different ``types'' of possible likely competitors that can emerge \emph{ex post}, and prize differentials at different ranks can be leveraged to create incentives for competition with a particular type. It is also clear from (\ref{g_general}) that for the EPS prize schedule, when the only positive prize differential is $v_n-v_{n+1}=\frac{1}{n}$, the global mode is always optimal.

Since the lowest possible optimal standard is at the global mode, it is of interest to consider cases where, despite multimodality, the standard at the global mode is optimal for any ${\bf v}\in\mathcal{V}$. The result is as follows.

\begin{proposition}
\label{prop_one_standard}
For any prize schedule $\mathbf{v}\in\mathcal{V}$ it is optimal to set the standard at the global mode, $\rho^*=e^*+x_m$, with $e^*=c'^{-1}(g(x_m;\mathbf{v}))$, if and and only if it is optimal to do so for the WTA prize schedule or, equivalently, $B_1(t)$ is maximized at the global mode $x_m$.
\end{proposition}

Note from (\ref{g_general}) that $B_1(t)=g(t;\mathbf{v}^{\rm WTA})$, i.e., $B_1(t)$ is the marginal benefit of effort in the WTA tournament with standard $\rho=e^*+t$. The sufficiency part follows from the property that $\arg\max_tB_r(t)$ decreases with $r$ (see the proof in Appendix \ref{app_proofs} for details);\footnote{Considering prize schedules of the form ${\bf v}^{(s)}=(\frac{1}{s},\ldots,\frac{1}{s},0,\ldots,0)$ awarding $s\le n$ equal prizes at the top and $n-s$ zero prizes at the bottom, this property also implies that the optimal threshold level of noise, $t^*$, weakly decreases with $s$. That is, the optimal standard shifts to a weakly lower mode as prizes become more equal. Note that optimal prizes belong to this class, as shown in Section \ref{sec:opt_tournament}. For illustration, consider the red distribution in Figure \ref{fig:example_main}: The optimal threshold is at $t^*=1$ for ${\bf v}^{(1)}$ (WTA) and at $t^*=0.5$ for ${\bf v}^{(2)}$ (two equal prizes) and ${\bf v}^{(3)}$ (EPS).} and necessity obtains immediately by considering the WTA prize schedule. Intuitively, under the WTA prize schedule only the incentives at the very top matter, and hence, the optimal standard is the highest as compared to that under any other prize schedule. If, nevertheless, the optimal standard under the WTA is at the global mode, the optimal standard under any other prize schedule cannot be higher and is thus at the global mode too. This occurs when the modes to the right of the global one are small relative to it; for example, when the subpopulations of noise corresponding to different modes to the right of the global mode are not too large. Proposition \ref{prop_one_standard} is illustrated in Figure \ref{fig:example_main} where for the distribution shown in green, $B_1(t)=g(t;\mathbf{v}^{\rm WTA})$ is maximized at the global mode $x_m=0.5$, and functions $g(t;\mathbf{v})$ for the prize schedules with two and three equal prizes are also maximized at the global mode. For the red distribution, $B_1(t)$ is maximized at the mode $x=1$, i.e., above $x_m$, while  functions $g(t;\mathbf{v})$ for the prize schedules with two and three equal prizes are maximized at the global mode $x_m$. Comparing the densities of the green and the red distributions (top left panel of Figure \ref{fig:example_main}) one can see that while the densities at the global mode $x_m=0.5$ are almost the same, the density of the green distribution at the mode $x=1$ is much smaller than that of the red distribution.

Finally, observe that Propositions \ref{prop_opt_reserve} and \ref{prop_one_standard} hold if players are risk averse, with a separable utility over money and effort, and $v_r$ is interpreted as the utility of receiving the $r$-th prize.

\subsection{Optimal Prize Schedules}
\label{sec:opt_tournament}

The structure of the marginal benefit function $g(t;{\bf v})$ in (\ref{g_general}) is such that the marginal probabilities of gaining different prize differentials are multiplied by the prize differentials themselves; therefore, in addition to setting the standard the principal can also affect incentives by choosing a prize schedule ${\bf v}$.

Since, according to Proposition \ref{prop_opt_reserve}, the \emph{optimal} standard only depends on ${\bf v}$ through the implemented equilibrium effort $e^*$, we can fully characterize optimal tournament-with-a-standard pay schemes where the principal can choose both $\rho$ and ${\bf v}$.

While in general the characterization is complex and sensitive to the shape of $f$, it is simplified substantially in two prominent special cases. Let $h(x)=\frac{f(x)}{1-F(x)}$ denote the hazard (or failure) rate of noise. A distribution (or the corresponding random variable) has an \emph{increasing failure rate} (IFR) if $h(\cdot)$ is increasing, and a \emph{decreasing failure rate} (DFR) if $h(\cdot)$ is decreasing, in the support. The IFR family includes many standard distributions such as normal, Gumbel, or logistic. All log-concave distributions are IFR, but not the other way around. DFR distributions necessarily have decreasing densities and include, but are not limited to, log-convex distributions with $f(\overline{x})=0$, such as Pareto. The exponential distribution with $F(x)=1-\exp(-\lambda x)$ has a constant failure rate (equal to $\lambda$) and separates the two families. Therefore, IFR distributions are often referred to as having sub-exponential or \emph{light} tails, whereas DFR distributions are \emph{heavy}-tailed. The following proposition provides a sharp characterization of optimal tournament-with-a-standard pay schemes. 

\begin{proposition}
\label{prop_opt_tournament_with_reserve}
(i) Suppose $f$ is such that the condition in Proposition \ref{prop_one_standard} is satisfied, with global mode $x_m$, and IFR for $x>x_m$. Then the WTA prize schedule and a standard at $\rho^*=e^*+x_m$, where $e^*=c'^{-1}(g(x_m;{\bf v}^{\rm WTA}))$, maximize the equilibrium effort.

(ii) Suppose $f$ is DFR. Then the EPS prize schedule and a standard at $\rho^*=e^*+\underline{x}$, where $e^*=c'^{-1}(\frac{f(\underline{x})}{n})$, maximize the equilibrium effort.

(iii) For the exponential distribution, $F(x)=1-\exp(-\lambda x)$, $x\ge 0$, any prize schedule ${\bf v}\in\mathcal{V}$ and a standard at $\rho^*=e^*=c'^{-1}(\frac{\lambda}{n})$, maximize the equilibrium effort.
\end{proposition}
Part (i) of Proposition \ref{prop_opt_tournament_with_reserve} is illustrated in Figure \ref{fig:example_main} where the WTA prize schedule produces the largest function $g(t;{\bf v})$ for any $t$, for all three distributions. Importantly, this distribution is not unimodal and is not IFR globally; yet, it satisfies the weaker conditions of part (i) that $f$ is IFR to the right of the global mode (see the top left panel showing the density and the bottom left panel showing the hazard rate).

\begin{figure}[tbp]
\centering
{%
\includegraphics[width=3in]{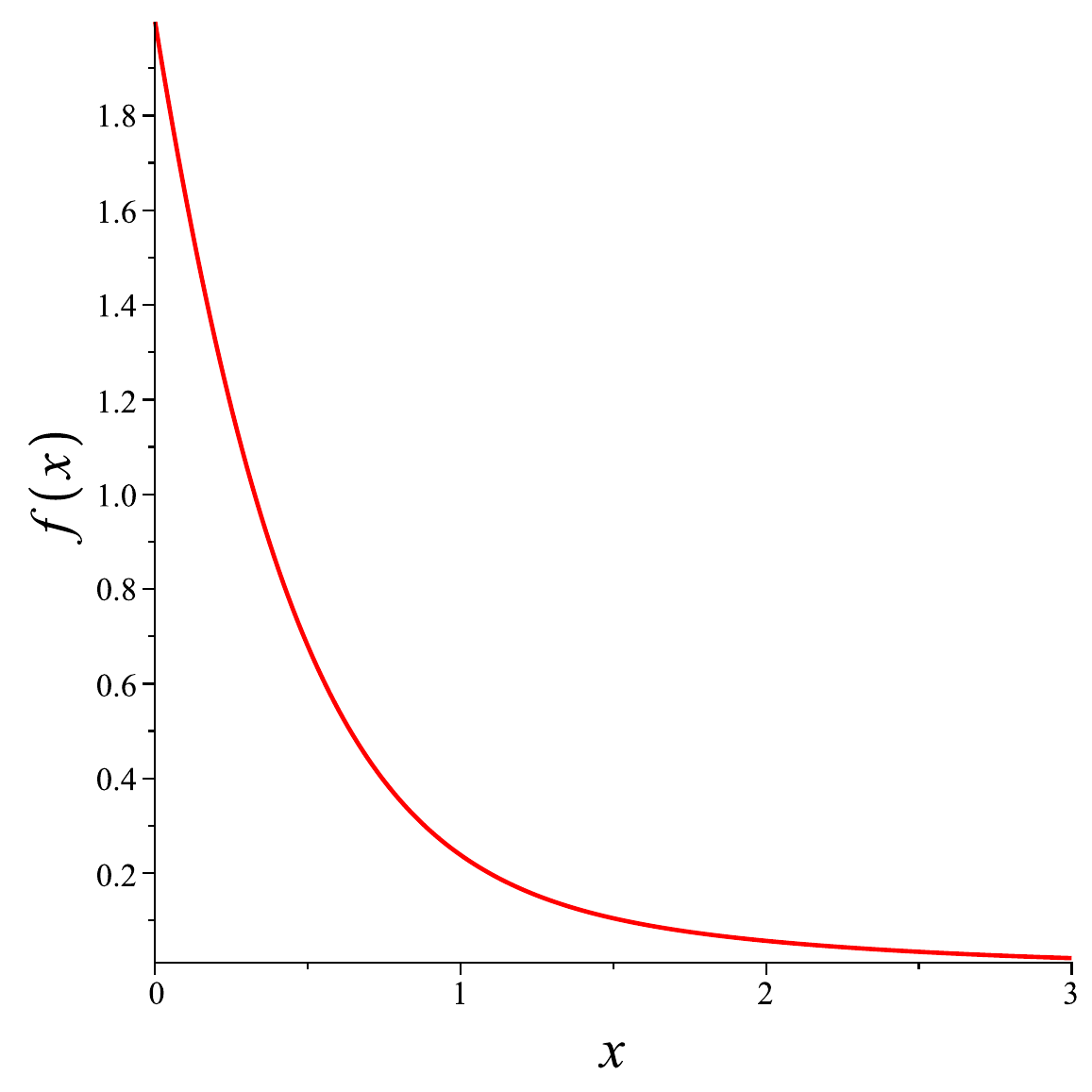}
}
\hfill
{%
\includegraphics[width=3in]{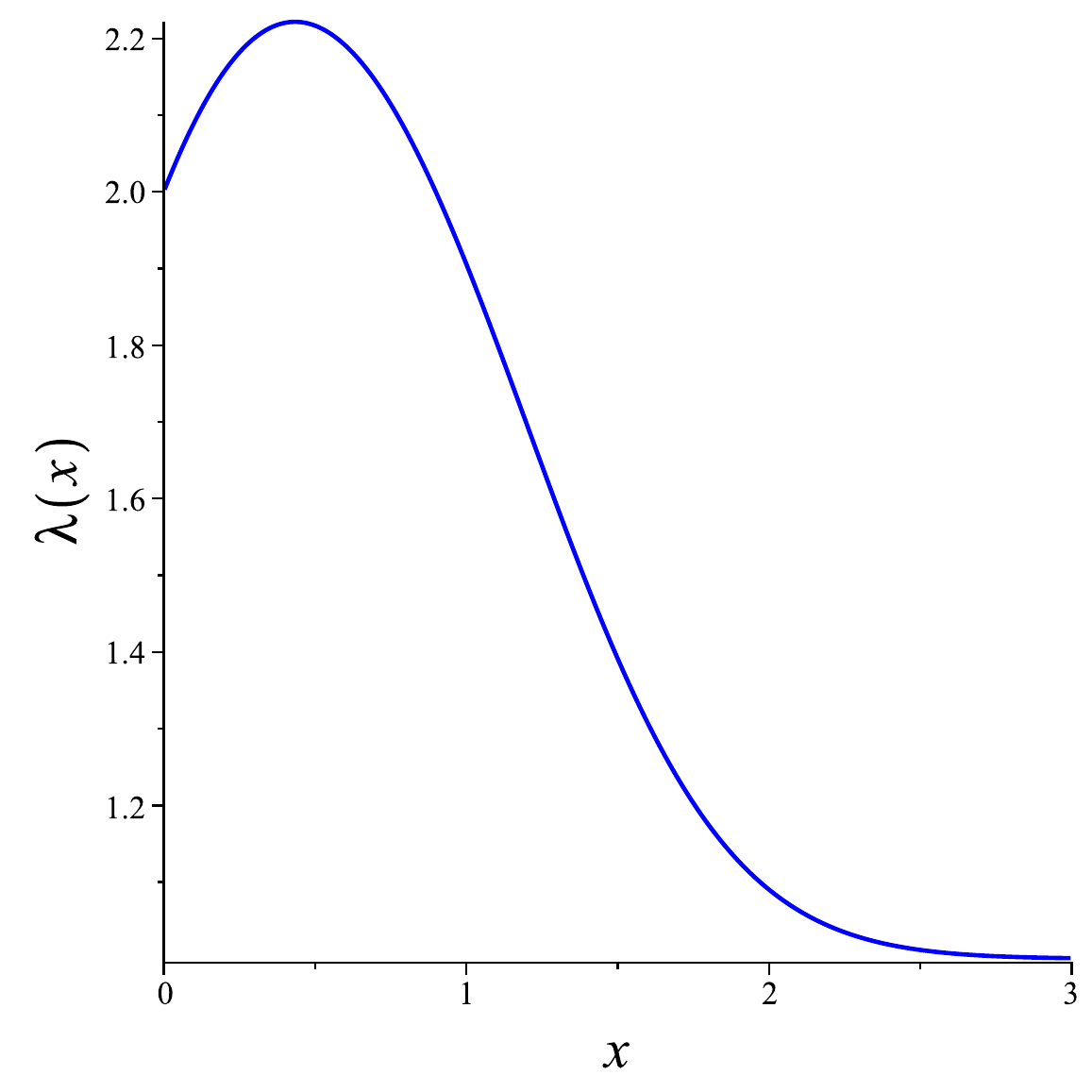}
}
{%
\includegraphics[width=3in]{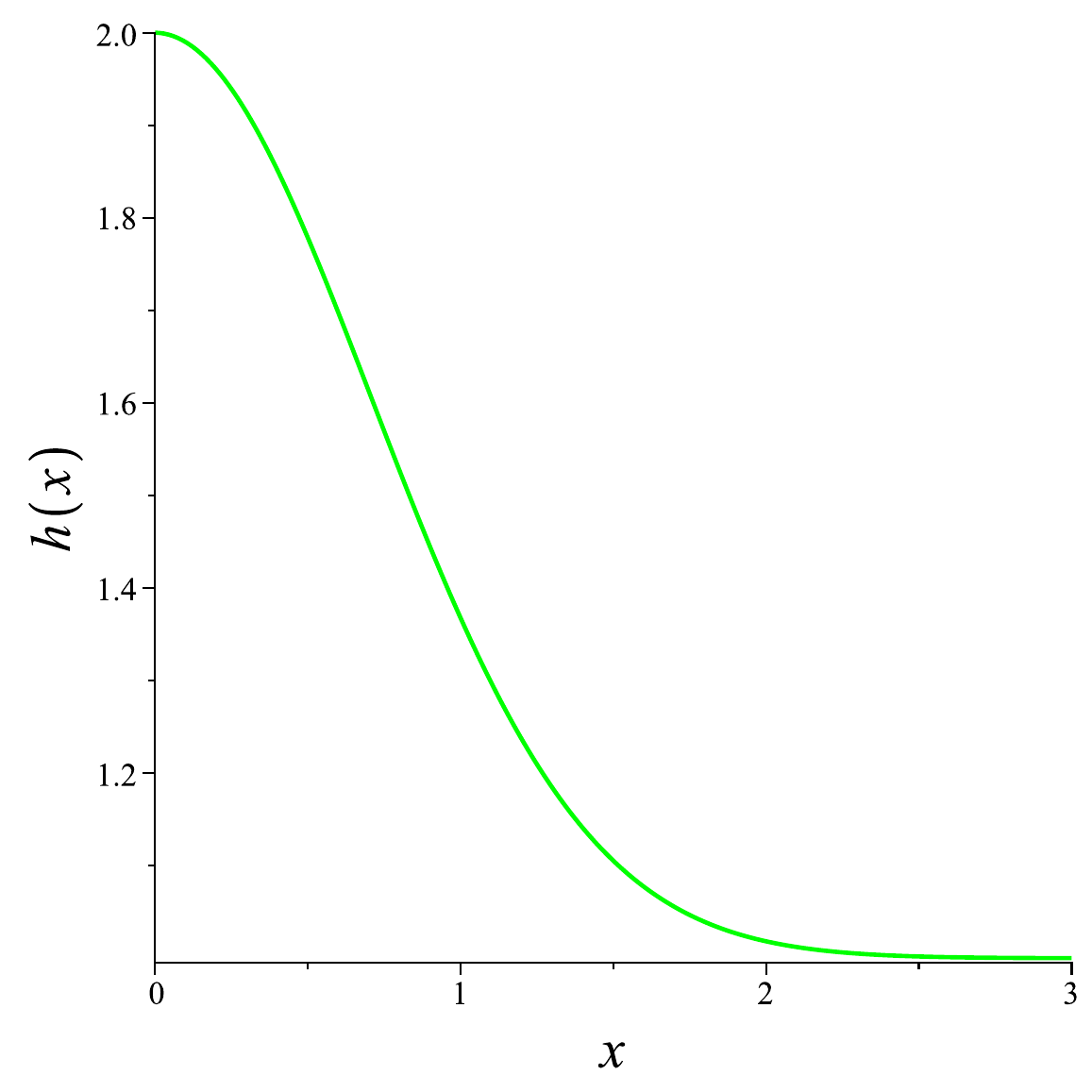}
}
\hspace{10pt}
{%
\includegraphics[width=3in]{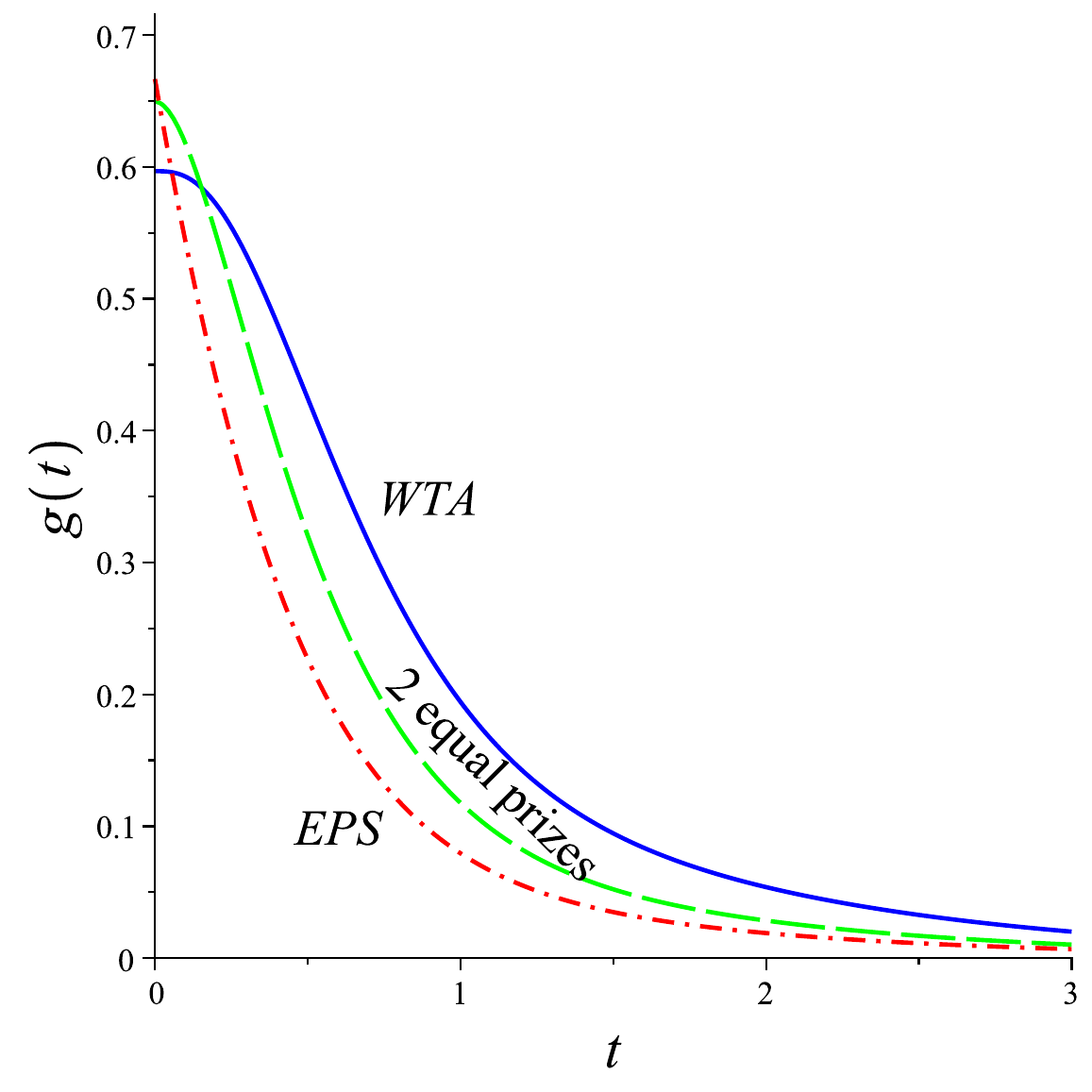}
}
\caption{\small{The density $f$ (top left), likelihood ratio $\lambda=-f'/f$ (top right), hazard rate $h=f/(1-F)$ (bottom left) and function $g(t;{\bf v})$ (bottom right, for different prize schedules ${\bf v}$) for a DFR distribution $F(x)=1-\exp\left[-x-\frac{\sqrt{\pi}}{2}\mathrm{erf}(x)\right]$, $x\ge 0$, whose hazard rate is $h(x)=1+\exp(-x^2)$. Parameters: $n=3$; ${\bf v}=(1,0,0)$ for WTA,  $(1/2,1/2,0)$ for two equal prizes, and $(1/3,1/3,1/3)$ for EPS.}}
\label{fig:example_DFR}
\end{figure}

Part (ii) of Proposition \ref{prop_opt_tournament_with_reserve} is illustrated in Figure \ref{fig:example_DFR} with a DFR distribution $F(x)=1-\exp\left[-x-\frac{\sqrt{\pi}}{2}\mathrm{erf}(x)\right]$, $x\ge 0$, whose hazard rate is $h(x)=1+\exp(-x^2)$ and $x_m=0$. As seen from the bottom right panel, $g(0;{\bf v})$ is maximized by the EPS prize schedule. Note that in the equilibrium all the players pass the standard and get the prize $\frac{1}{n}$.  Effectively, there is no competition between the players, and the tournament can be replaced by individual bonuses which are paid to all who qualify. This outcome can also be replicated by a lottery which awards the whole prize budget to one of the players who passed the standard---which is useful if the prize budget is indivisible.

To understand Proposition \ref{prop_opt_tournament_with_reserve}, consider (\ref{g_general}) with $t=x_m$. Optimal prizes solve the linear programming problem $\max_{{\bf v}\in\mathcal{V}}g(x_m;{\bf v})$ that, by introducing prize differentials $d_r=v_r-v_{r+1}$, can be transformed into
\begin{align}
\max_{d_1,\ldots,d_n\ge 0} \sum_{r=1}^nB_r(x_m)d_r \quad {\rm s.t.\,} \sum_{r=1}^nrd_r=1,
\end{align}
where the last part is the transformed budget constraint $\sum_{r=1}^nv_r=1$. This problem has the form of a utility maximization problem where coefficients $B_r(x_m)$ represent the (constant) marginal ``utilities'' of each prize differential $d_r$, and the ``price'' of the $r$-th prize differential is $r$. Therefore, generically, the optimal prize allocation is a corner solution such that $d_{r^*}>0$ for some $r^*$ and $d_r=0$ for all $r\ne r^*$, where $r^*$ maximizes $\frac{B_r(x_m)}{r}$. Such a solution corresponds to a prize schedule ${\bf v}=(\frac{1}{r^*},\ldots,\frac{1}{r^*},0,\ldots,0)$ with $r^*$ equal, positive prizes at the top and $n-r^*$ zero prizes at the bottom. Proposition \ref{prop_opt_tournament_with_reserve} is based on the representation
\begin{align}
\label{B_hazard}
\frac{B_r(x_m)}{r} = \frac{1}{n}\int \tilde{h}(x;x_m)\dd F_{(n-r:n)}(x),
\end{align}
which is obtained from (\ref{B_r_f_tilde}) (see the proof of Proposition \ref{prop_opt_tournament_with_reserve} for details). Here, $\tilde{h}(x;t) = \frac{\tilde{f}(x;t)}{1-F(x)}$ is a modified hazard rate of noise based on function $\tilde{f}(x;t) = f(\max\{x,t\})$. For distributions that are IFR for $x>t$, $\tilde{h}(x;t)$ is increasing in $x$ for any $t$,\footnote{\label{fn. WTA for any standard}This implies, in particular, that for distributions that are IFR for $x>t$ the WTA prize schedule maximizes effort, for a given standard, even if $f$ does not satisfy the condition in Proposition \ref{prop_one_standard} and the standard is not optimally set.} whereas for DFR distributions $\tilde{h}(x;t)$ is decreasing when $t=\underline{x}$---the lower bound of the support of $F$, which incidentally also equals $x_m$ because the density is always decreasing in the DFR case. The three parts of Proposition \ref{prop_opt_tournament_with_reserve} then follow immediately from (\ref{B_hazard}).

Figure \ref{fig:example_main} also confirms that for IFR distributions of noise the WTA prize schedule is optimal for \emph{any} standard above which the distribution is IFR (see fn. \ref{fn. WTA for any standard}). At the same time, Figure \ref{fig:example_DFR} shows that when noise is DFR the EPS schedule may or may not be optimal if the standard is not at the mode.\footnote{See the end of Section \ref{sec:opt_cardinal} for a discussion of the optimal prize schedule in this case.} This is indeed the case in general. It is clear from (\ref{B_hazard}) that $\frac{B_r(t)}{r}$ is decreasing in $r$ for any $t$ when $f$ is IFR; it is not, however, increasing in $r$ for any $t$ when $f$ is DFR.

Finally, the problem of choosing the optimal prize schedule in tournaments \emph{without a standard} was solved by \cite{Drugov-Ryvkin:2020} who show that the optimal prize allocation is WTA if $f$ is IFR and punish-the-bottom (PTB), i.e., awarding equal prizes to all ranks except $r=n$, if $f$ is DFR. The characterization in Proposition \ref{prop_opt_tournament_with_reserve} differs in two important ways. First, for WTA to be optimal with the optimal standard, we only require for $f$ to be IFR for $x>x_m$; that is, $f$ does not need to be IFR globally. For example, the three distributions in Figure \ref{fig:example_main}
are not IFR globally. Moreover, even if a standard is not chosen optimally, it is sufficient for $f$ to be IFR for $x>t$ for WTA to be optimal (see fn. \ref{fn. WTA for any standard}). Second, the EPS schedule is optimal when $f$ is DFR and the standard is at the lower bound of the equilibrium distribution of output. Without a standard, the EPS schedule would lead to zero effort, and the PTB schedule is the next-best solution in the DFR case. However, if the principal can impose a standard, the optimal pay scheme is such that in equilibrium all players pass and there is no competition.

\section{Optimal Cardinal Pay Schemes}
\label{sec:opt_cardinal}

The optimal tournaments with a standard identified in Proposition \ref{prop_opt_tournament_with_reserve} rely on the properties of the hazard rate of noise. In this section, we show that under stronger assumptions these pay schemes are also optimal in a much wider class of pay schemes where compensation can be conditioned on players' cardinal outputs. Let ${\bf w}:\mathds{R}^n\to\mathds{R}_+^n$ denote a general cardinal pay scheme in which $w_i({\bf y})$---player $i$'s compensation---may depend on the entire vector of outputs ${\bf y}=(y_1,\ldots,y_n)$. We consider pay schemes satisfying the following properties for all ${\bf y}\in\mathds{R}^n$ and $i\in\{1,\ldots,n\}$.

\begin{assumption}
\label{ass_w}
(a) Anonymity: $w_i(y_1,\ldots,y_n) = w_{\tau(i)}(y_{\tau(1)},\ldots,y_{\tau(n)})$ for any permutation $\tau$ of indices $\{1,\ldots,n\}$;

(b) Monotonicity: $w_i({\bf y})$ is increasing in $y_i$;

(c) Budget constraint: $\sum_{i=1}^nw_i({\bf y})\le 1$.
\end{assumption}

\noindent Assumption (a) ensures that players are treated symmetrically and would not permit situations, for instance, where some players are excluded from the tournament. While in some cases it can be optimal to restrict the number of players (for example, when total effort decreases with $n$), the symmetry assumption is normatively attractive when players are symmetric \emph{ex ante}.\footnote{Our results still apply to quasi-symmetric schemes where some players are excluded and the rest are treated symmetrically; however, they do not apply when active players can be treated asymmetrically.} Assumption (b) is standard in the moral hazard literature, with the argument being that its violation can create an incentive for players to destroy output. Finally, assumption (c) is consistent with the budget constraint we imposed for tournament schemes. The next proposition is our main result in this section.

\begin{proposition}
\label{prop_general_w}
The optimal tournaments with a standard---WTA and EPS with standards at the modal performance---are optimal in the class of pay schemes satisfying Assumption \ref{ass_w} for $f$ log-concave and log-convex, respectively.
\end{proposition}

Proposition \ref{prop_general_w} shows that when the noise is log-concave or log-convex, the restriction to tournaments with a standard is without loss of generality if one considers arbitrary pay schemes based on fully contractible cardinal outputs, including individual contracts as well as arbitrarily complex relative performance schemes. An obvious advantage of tournament schemes is in that they rely only on ordinal output comparisons. Our proof of Proposition \ref{prop_general_w} is based on establishing an upper bound on effort in general contracts satisfying Assumption \ref{ass_w} and then showing that the two optimal pay schemes achieve the corresponding upper bounds in the two cases.\footnote{It is also possible to prove Proposition \ref{prop_general_w} by treating players' individual noise realizations as their \textit{ex post} types and using the approach of \cite{Myerson:1981} (we thank Alex Suzdaltsev for pointing this out in personal communication).} The equilibrium existence conditions underlying Proposition \ref{prop_general_w} are similar to those for Propositions \ref{prop_opt_reserve} and \ref{prop_opt_tournament_with_reserve} (see fn. \ref{fn eq existence}).

Proposition \ref{prop_general_w} implies that, when noise is log-concave or log-convex, \emph{conditional} prize schedules do not lead to a higher effort. These are the prize schedules in which the principal can commit to awarding a different vector of prizes ${\bf v}^s=(v_1^s,\ldots,v^s_s)$ for each number of players $s=1,\ldots,n$ passing the standard. They are more general than the \emph{unconditional} prize schedules we considered in this paper, but of course are still ordinal and hence, less general than the cardinal pay schemes studied in this section. However, when the noise is neither log-concave nor log-convex and/or the standard is not set optimally, such prize schedules might generate strictly higher efforts than unconditional ones. For example, as we discussed after Proposition \ref{prop_opt_tournament_with_reserve}, when the noise is DFR and the standard is not set at the optimal level, the EPS schedule might not be optimal (see the bottom right panel in Figure \ref{fig:example_DFR}). For log-convex noise distributions, it can be shown that the \emph{conditional} EPS schedule, in which the prize budget is divided equally among the players who pass the standard (i.e., $v_r^s=\frac{1}{s}$ for all $r=1,\ldots,s$), is optimal for any standard. %This result is not in our previous versions, but it was something I've shown later. I doubt it holds for DFR, but I didn't check.
When the standard is optimally set at the lower bound of the support of performance, $\rho^*=e^*+\underline{x}$, it is passed with probability one in equilibrium. Therefore, only $s=n$ matters in the equilibrium and the unconditional EPS schedule generates the same effort as the conditional one. But for any other standard the two schemes lead to different equilibrium efforts.\footnote{See the extended working paper \cite{Drugov-Ryvkin-Zhang:2022} for more results on conditional prize schedules.}

Recall that log-concave distributions are IFR and log-convex distributions with $f(\overline{x})=0$ are DFR, but not the other way around. That is, Proposition \ref{prop_general_w} gives the optimality of the WTA and EPS schemes, with the corresponding standards, for a smaller class of noise distributions as compared to Proposition \ref{prop_opt_tournament_with_reserve}. The reason is that, as discussed in the next section, log-concave and log-convex distributions have monotone likelihood ratios. The results of this section can, therefore, be obtained using insights from the moral hazard literature relying on the MLRP assumption. At the same time, IFR and DFR distributions do not have monotone, or even single-crossing, likelihood ratios in general. Consider, for example, the likelihood ratios in Figure \ref{fig:example_main} for $x \geq 0.25$. For distributions that do not satisfy MLRP, the optimality of the two tournament schemes in the class of cardinal schemes cannot generally be established.

\paragraph{Relation to the literature on moral hazard} 
The rank-order tournament model with additive noise can be viewed as a special case of a more general model of stochastic performance used in the literature on moral hazard \citep[for a textbook exposition see, e.g.,][]{Laffont-Martimort:2002}, where it is assumed that agent $i$'s output $Y_i$ is distributed according to some $Q(y|e_i)$, with density $q(y|e_i)$, that is stochastically increasing in $e_i$. The most important assumption about this distribution is the monotone likelihood ratio property (MLRP) whereby the likelihood ratio $L(y|e)=\frac{q_e(y|e)}{q(y|e)}$ is increasing in $y$.\footnote{Another assumption is the convexity of the distribution function condition (CDFC); however, it is purely technical and ensures that the agent's problem is globally concave for any convex effort cost function. We do not impose the CDFC and instead follow the approach of the tournament literature starting with \cite{Lazear-Rosen:1981} and require sufficient convexity of the cost function, see fn. \ref{fn eq existence}. Note also that for additive noise the CDFC is equivalent to density $f$ being monotone increasing, which contradicts our assumption that $f(\overline{x})=0$. Moreover, in this case the CDFC implies that the likelihood ratio $L$ is everywhere negative.} It ensures that optimal contracts are monotone in output.
In other words, if it is posited that pay schemes \emph{must} be monotone, this constraint is not binding in the principal's problem under the MLRP. Under these conditions, it has been shown \citep{Holmstrom:1979} that the shape of optimal rewards follows that of $L$; in particular, under limited liability, only output levels such that $L\ge 0$ should be rewarded.

\cite{Kirkegaard:2023} considers contest design in a general moral hazard framework with heterogeneous agents. He assumes the MLRP to ensure that the monotonicity constraint on rewards is not binding. The key insight of \cite{Kirkegaard:2023} is that the role of likelihood ratio in this multi-agent setting is similar to that in the standard moral hazard problem; namely, that only positive likelihood ratios should be rewarded. In the case of additive noise, the likelihood ratio is $L(y|e) = -\frac{f'(y-e)}{f(y-e)}$, and hence as long as $f$ is either (i) unimodal or (ii) monotonically decreasing, the optimal cutoff point is at the mode of $f$, consistent with the location of the optimal standard at $e^*+x_m$ in our Proposition \ref{prop_opt_reserve}. However, the intuition behind this result in our model is different. As seen from our discussion of Proposition \ref{prop_opt_reserve}, in our setting the location of the optimal standard at the mode is not related to the likelihood ratio being positive to the right of the mode; rather it is related to the density of $X-\max\{X_{(n-r:n-1)},t\}$ at zero being maximized by $t=x_m$ even when the likelihood ratio is not single-crossing (see Figure \ref{fig:example_main}).

With respect to prize allocations, for a given standard, the key insight from \cite{Kirkegaard:2023} is again that higher likelihood ratios should be rewarded more. Thus, under the MLRP (or, in our case, log-concave noise) it is optimal to award the entire prize to the top performer, producing the WTA schedule. For log-convex noise, the likelihood ratio is decreasing (and positive everywhere); thus, unrestricted optimal prizes would decline with rank, which is impossible, and hence the next-best prize schedule is EPS. However, our Proposition \ref{prop_opt_tournament_with_reserve} shows that the optimality of WTA does not rely on the likelihood ratio being increasing or even single-crossing. A much weaker condition---that $f$ is IFR above the standard---is sufficient. Similarly, the optimality of EPS does not rely on the likelihood ratio being decreasing in the DFR case (see Figure \ref{fig:example_DFR}).

As expected, since tournaments with a standard represent a restricted class of contracts, their optimality is controlled by different forces and can be characterized under weaker conditions as compared to more general cardinal contracts. Under stricter conditions when the insights of \cite{Kirkegaard:2023} apply---for log-concave and, by extention, log-convex noise---our Proposition \ref{prop_general_w} shows that the restriction to tournaments with a standard is without loss of generality.

\section{Performance Standards in Other Contest Models}
\label{sec:applications}

In this section, we consider three alternative models of noisy contests---Tullock contests \citep{Tullock:1980}, innovation contests  \citep{Fullerton-McAfee:1999}, and patent races \citep{Loury:1979}. These models are closely related to each other \citep{Baye-Hoppe:2003}, and can be derived from the Lazear-Rosen model using a specific distribution of noise  \citep{Fu-Lu:2012}. Our results, therefore, can be applied to characterize the optimal standard and optimal pay scheme in each of these models.

\subsection{Tullock Contests with a Standard}

The \cite{Tullock:1980} model of winner-take-all contests is arguably the most popular one in the literature. The corresponding contest success function (CSF) 
\begin{equation}
\label{CSF_Tullock}
p_i({\bf e})= \frac{e_i}{\sum_{j=1}^ne_j}
\end{equation}
is player $i$'s probability of winning the contest for a given vector ${\bf e}=(e_1,\ldots,e_n)$ of players' efforts. It is well known---going back at least to the logit analysis of \cite{McFadden:1974}---that (\ref{CSF_Tullock}) can be derived from a Lazear-Rosen model with performance $\hat{Y}_i=\hat{e}_i+\hat{X}_i$ and $\hat{X}_i$ following the Gumbel distribution, by taking the exponential transformation $e_i=\exp(\hat{e}_i)$. Using this approach when there is a standard $\hat{\rho}$ yields the CSF
\begin{equation}
\label{CSF_Tullock_ours}
\tilde{p}_i({\bf e};\rho) = \frac{e_i}{\sum_{j=1}^ne_j}\left[1-\exp\left(-\frac{1}{\rho}\sum_{j=1}^ne_j\right)\right],
\end{equation}
with $\rho=\exp{\hat{\rho}}$. As expected, $\tilde{p}_i({\bf e};\rho)$ decreases with $\rho$ and converges to $p_i({\bf e})$ as $\rho\to 0$ (or $\hat{\rho}\to-\infty$ in the additive representation).

By Proposition \ref{prop_opt_reserve}, the optimal standard is  $\hat{\rho}^*=\hat{e}^*+\hat{x}_m$. Since $\hat{x}_m=0$ for the Gumbel distribution, applying the exponential transformation we get $\rho^*=e^* \exp{(0)}=e^*$. It can then be shown from (\ref{g_general}) that, for a linear cost function $c(e)=e$, the equilibrium effort becomes
\begin{equation}
\label{Tullock_effort_opt}
e^* = \frac{n-1}{n^2} + \frac{1}{n^2}\exp(-n).
\end{equation}
The first term is the same as in the usual Tullock contest without a standard, and the second term emerges due to the optimal choice of the standard.

Note that the CSF in (\ref{CSF_Tullock_ours}) also describes a contest with a possibility of ties. Indeed, with probability $\exp\left(-\frac{1}{\rho}\sum_{j=1}^ne_j\right)$ there is no winner and the prize is not awarded. \cite{Nalebuff-Stiglitz:1983} were arguably the first to suggest that a requirement to ``win by a gap'' in a Lazear-Rosen tournament---another way to model a tie---may raise effort \citep[for a recent review of that literature see, e.g.,][]{Vesperoni-Yildizparlak:2019}. Proposition \ref{prop_general_w}, however, implies that any gap leads to a lower equilibrium effort than the optimal standard.

\subsection{Innovation Contests with a Standard}
\label{sec:FMA}

\cite{Fullerton-McAfee:1999} consider an innovation tournament model in which the distribution of player $i$'s performance is given by $H(x)^{e_i}$, where $H$ is an arbitrary absolutely continuous distribution with support $[0,\bar{x}]$, and $e_i$ is the player's effort. The interpretation is that players can draw ideas from $H$ and use the best draw, and $e_i$ represents the number of draws. The resulting CSF then takes the form (\ref{CSF_Tullock}). A standard in this model can be interpreted as a pre-existing level of knowledge $\rho\in[0,\bar{x}]$ such that ideas below $\rho$ do not constitute an innovation. It can also be a minimum standard set by the principal. For example, in the case of the Netflix prize mentioned in the Introduction, Netflix had its own proprietary algorithm and expected the contest participants to sufficiently improve on it. The resulting CSF with a standard,
\begin{align}
\label{CSF_FM_reserve}
\tilde{p}_i({\bf e};\rho) = \frac{e_i}{\sum_{j=1}^ne_j}\left[1-H(\rho)^{\sum_{j=1}^ne_j}\right],
\end{align}
has the same structure as (\ref{CSF_Tullock_ours}).\footnote{In fact, it coincides with (\ref{CSF_Tullock_ours}) if $H(x)=\exp(-\frac{1}{x})$, i.e., if ideas follow the inverse exponential distribution. This is to be expected, since the Gumbel distribution underlying the Tullock CSF is obtained from the inverse exponential distribution via exponentiation.} Recall that the optimal standard in the Tullock model is $\rho^*=e^*$. Comparing (\ref{CSF_FM_reserve}) with (\ref{CSF_Tullock_ours}), we, therefore, obtain that the optimal standard in the model of \cite{Fullerton-McAfee:1999} is $\rho^* = H^{-1}(\exp(-\frac{1}{e^*}))$, where $e^*$ is the symmetric equilibrium effort given by (\ref{Tullock_effort_opt}). See also \cite{Kirkegaard:2023} for analysis of the optimal standard in a more general version of this model.

\subsection{Patent Races with a Deadline}

\cite{Loury:1979} and \cite{Dasgupta-Stiglitz:1980} initiated the literature on patent races. Suppose there are $n$ firms in the market which compete by investing in R\&D. Each firm's success follows random arrival with an exponentially distributed time, where the parameter of the exponential distribution, i.e., the expected time to innovation, is decreasing in the firm's investment. The probability for a firm to win the race is then given by the Tullock CSF (\ref{CSF_Tullock}), and \citet[][Equation (6)]{Loury:1979} obtains a CSF of the form (\ref{CSF_Tullock_ours}) as an intermediate step for a firm that tries to beat the lowest arrival time among its rivals.

One can think, more generally, of a race with an exogenous \emph{deadline}, $\tau>0$, such that innovations arriving after $\tau$ are not rewarded. To apply our results, suppose firm $i$'s innovation arrives at a random time $T_i=\exp(-\hat{e}_i-\hat{X}_i)$, where $e_i=\exp(\hat{e}_i)$ is the firm's innovation effort (broadly defined), and $\hat{X}_i$ is a shock. The probability for a firm to win the race is given by the Tullock CSF (\ref{CSF_Tullock}) if $\hat{X}_i$ has the Gumbel distribution. CSF (\ref{CSF_Tullock_ours}) describes this patent race with standard $\rho=\frac{1}{\tau}$, and hence $\tau^*=\frac{1}{e^*}$ is the optimal deadline. Other distributions of $\hat{X}_i$ will lead to different CSFs, the optimal deadline $\tau^*=\frac{1}{e^*}\exp(-\hat{x}_m)$, where $\hat{x}_m$ is a mode of $\hat{X}_i$, and a possibility that it is optimal to award more than one prize in the race.

\section{Conclusion}
\label{sec:conclusion}

This paper is the first to study the role of performance standards in rank-order tournaments. Standards play an important role in tournament design. In various settings with competitive incentives, such as the assignment of raises, bonuses or promotions in organizations, or innovation contests, it is in the principal's interest to not only reward the best performer(s) but also to ensure that the performance of those selected exceeds a minimum standard. We show that the optimal standard is always at a mode of the performance distribution, and provide a condition for when it is at the global mode. We find optimal prize schemes---to be used together with the optimal standard---for distributions with monotone hazard rates and also show that when the noise distribution is log-concave or log-convex, a tournament with the optimal standard and prize schedule achieves optimality in a much wider class of monotone, symmetric pay schemes based on cardinal performance.

%\textcolor{cyan}{I am not sure about this paragraph since the empirical applications that follow are about unimodal distributions. Keeping it will make the empirical stuff look really underdeveloped (even more that what it is now).} Settings with light-tailed and heavy-tailed shocks can be broadly characterized as those where luck plays a relatively smaller and larger role, respectively, in tournament outcomes. Our results imply that it is not a good idea to use tournaments under heavy-tailed shocks, at least if the goal is to motivate effort. A better approach is to impose a mild performance standard and use individual bonuses.

Let us finish by discussing how one can use our results to identify from data whether or not a standard is chosen optimally in a tournament. With a large enough performance dataset, it is straightforward to estimate the distribution of performance and its modes nonparametrically. Suppose, for simplicity, that it is unimodal. %\textcolor{red}{I am not sure it's a good idea to say it is the most common case because then the question becomes, why we are making such a big deal about the multimodal case.}
The recommendation then is to raise (respectively, lower) the standard if it is below (respectively, above) the modal performance. Moreover, normalizing the mean of noise to zero, the symmetric equilibrium effort can be estimated as the mean of performance, and then we recover the distribution of shocks. If accounting information on costs is also available, it is then possible to estimate the optimal standard. 

We can also assess whether a standard is chosen optimally by looking at the fraction of players passing the standard. Indeed, the equilibrium probability of passing the optimal standard in our model, $1-F(x_m)$, is 50\% if the (unimodal) distribution of noise is symmetric and greater than 50\% if the distribution is positively skewed. Note also that the probability of passing the optimal standard is independent of the prize scheme or the number of players. Therefore, if in similar tournaments with different numbers of players the fraction of passing players is different, it may indicate that at least in some of them the standard is not optimally set. Finally, suppose that the contest organizer has a higher budget than in a similar previous tournament and scales all the prizes up. This increases the equilibrium efforts and, since the noise distribution is unchanged, the performance standard should go up too.

%\newpage

\phantomsection
\addcontentsline{toc}{section}{References}
\bibliographystyle{aea}

\newpage
\appendix

\section{Proofs}
\label{app_proofs}

\begin{proof}{\bf of Proposition \protect\ref{prop_opt_reserve}}

\noindent Let $m_1>\ldots>m_K$ denote the modes of $f$.\footnote{Point $m\in\mathds{R}$ is called a \emph{mode} if there exists an interval $(a,b)\ni m$ such that $f$ is increasing and nonconstant in $(a,m]$ and decreasing and nonconstant in $[m,b)$. Modes $m<m'$ are \emph{distinct} if there exists an $x_0\in(m,m')$ such that $f(x_0)<\min\{f(m),f(m')\}$. We select the modes $m_1,\ldots,m_K$ to be distinct. The largest number of distinct modes is called the number of modes and denoted by $K$. We assume $K$ is finite.} We refer to the $k$-th element of this sequence as the $k$-th mode, and to the largest element of this sequence that maximizes $f$ globally as the \emph{global mode}. The index of the global mode is denoted $\hat{k}$. 

Equation (\ref{B_r}) can be written as (\ref{B_r_f_tilde}), with $\tilde{f}(x;t) = f(\max\{x,t\})$. Notice first that if $t$ is not at a mode of $f$, then one can select a mode $m_k$ such that $\tilde{f}(x;m_k)\ge \tilde{f}(x;t)$ for all $x$. Indeed, if $t<m_K$, select $m_K$; if $t>m_1$, select $m_1$; otherwise, if $t\in(m_{k+1},m_k)$ for some $k$ then either $m_k$ or $m_{k+1}$ can be selected. Thus, $g(t;{\bf v})$ is maximized by $t=m_k$ at some mode $m_k$. Second, for any mode $m_k<m_{\hat{k}}$ we have $\tilde{f}(x;m_{\hat{k}})\ge \tilde{f}(x;m_k)$ for all $x$. We conclude that $g(t;{\bf v})$ is maximized either at the global mode or at a mode to the right of the global mode.
\end{proof}

\bigskip

\begin{proof}{\bf of Proposition \protect\ref{prop_one_standard}}

\noindent {\bf Sufficiency}: Notice from (\ref{B_r_f_tilde}) that $B_r(t)$ cannot be maximized at a mode $m_k$ such that $f(m_l)>f(m_k)$ for another mode $m_l>m_k$ because $\tilde{f}(x;m_l)\ge \tilde{f}(x;m_k)$ would hold for all $x$, with strict inequality for a positive $X_{(n-r:n-1)}$-measure of $x$; therefore, all $B_r(t)$ are maximized at modes at or to the right of the global mode $m_{\hat{k}}$. Moreover, if there are any nonmonotonicities in the sequence $f(m_{\hat{k}}),\ldots,f(m_1)$ then the modes corresponding to local dips in this sequence cannot be optimal for any $B_r(t)$.

Consider now two adjacent, distinct modes $m<m'$ at or to the right of the global mode, where $f(m')<f(m)$, i.e., we skip any modes that correspond to local dips in the sequence above. Function $\tilde{f}(x;m)-\tilde{f}(x;m')$ is then single-crossing from positive to negative; therefore, the expectation of this function over $X_{(n-r:n-1)}$ decreases with $r$ decreasing, and hence the location of the maximum of $B_r(t)$ weakly increases with $r$ decreasing. Therefore, if $B_1(t)$ is maximized at $t=m_{\hat{k}}$ then all $B_r(t)$ are maximized at $t=m_{\hat{k}}$ as well.

\noindent {\bf Necessity}: Applying (\ref{g_general}) to the WTA prize schedule we have $g(t;1,0,\ldots,0) = B_1(t)$, and hence $B_1(t)$ has to be maximized at the global mode for the optimality of the global mode to hold for all prize schedules. 
\end{proof}

\bigskip

\begin{proof}{\bf of Proposition \protect\ref{prop_opt_tournament_with_reserve}}

\noindent To derive (\ref{B_hazard}), we introduce a modified hazard rate $\tilde{h}(x;t) = \frac{\tilde{f}(x;t)}{1-F(x)}$ and rewrite $B_r(t)$ in (\ref{B_r_f_tilde}) as follows:
\begin{align*}
& B_r(t) = \frac{(n-1)!}{(n-1-r)!(r-1)!}\int\tilde{f}(x;t)F(x)^{n-1-r}[1-F(x)]^{r-1}\dd F(x) \\
& = \frac{(n-1)!}{(n-1-r)!(r-1)!}\int \tilde{h}(x;t)F(x)^{n-1-r}[1-F(x)]^r\dd F(x)\\
& = \frac{r}{n}\frac{n!}{(n-1-r)!r!} \int \tilde{h}(x;t)F(x)^{n-1-r}[1-F(x)]^r\dd F(x)\\
& = \frac{r}{n}\int \tilde{h}(x;t)\dd F_{(n-r:n)}. 
\end{align*}

\noindent For part (ii), we have from (\ref{B_r}) and (\ref{g_general})
\[
g(\underline{x};{\bf v}^{\rm EPS}) = B_n(\underline{x})\frac{1}{n} = \frac{f(\underline{x})}{n}.
\]

\noindent For part (iii), note that the exponential density is decreasing and hence $x_m=\underline{x}=0$. Then $\tilde{h}(x;x_m)$ is a constant equal to $f(0)=\lambda$, any $r^*$ maximizes $\frac{B_r(x_m)}{r}$ in (\ref{B_hazard}), and the expression for $\rho^*=e^*$ follows from part (ii).
\end{proof}

\bigskip

\begin{proof}{\bf of Proposition \protect\ref{prop_general_w}}

\noindent In any symmetric pure strategy equilibrium where all players exert the same effort $e>0$, a symmetric FOC is satisfied. Without loss of generality, consider player 1 and assume that all players other than 1 choose effort $e$. Player 1's payoff from effort $e_1$ can be written as\footnote{The change of variable $x_1\rightarrowtail y_1=x_1+e_1$ is needed in order to shift the dependence on $e_1$ to the distribution of noise because $w_1$ may be discontinuous.}
\begin{align*}
&\pi_1(e_1,e) 
= \int \dd F(x_1)\ldots\int \dd F(x_n) w_1(e_1+x_1,e+x_2,\ldots,e+x_n) - c(e_1)\\
& = \int_{\underline{x}+e_1}^{\overline{x}+e_1}\dd y_1\int \dd F(x_2)\ldots \int \dd F(x_n) w_1(y_1,e+x_2,\ldots,e+x_n)f(y_1-e_1) - c(e_1).
\end{align*}
Next, we differentiate $\pi_1$ with respect to $e_1$, set $e_1=e$, and change the first variable of integration back to $x_1$. The result is
\begin{align}
\label{FOC_n_agents}
& \int \dd F(x_2)\ldots\int \dd F(x_n) \left[w_1(e+\overline{x},e+x_2,\ldots,e+x_n)f(\overline{x})-w_1(e+\underline{x},e+x_2,\ldots,e+x_n)f(\underline{x})\right]\nonumber\\
&-\int \dd x_1 \int \dd F(x_2)\ldots\int \dd F(x_n)w_1(e+x_1,\ldots,e+x_n)f'(x_1) = c'(e).
\end{align}

The left-hand side of (\ref{FOC_n_agents}) can be increased by dropping the negative term with $f(\underline{x})$ and shifting the lower limit of integration over $x_1$ to $x_m$. Using the assumption that $f(\overline{x})=0$, we obtain
\begin{equation}
\label{bound_n}
c'(e)\le R(e;{\bf w}) \equiv \int_{x_m}^{\overline{x}}\dd F(x_1)\int \dd F(x_2)\ldots\int \dd F(x_n) w_1(e+x_1,\ldots,e+x_n)\lambda(x_1).
\end{equation}
Here, $\lambda(x)=-\frac{f'(x)}{f(x)}$. The following lemma shows that function $R(e;{\bf w})$ can be uniformly bounded above by the expressions in the two cases of Proposition \ref{prop_opt_tournament_with_reserve}.

\begin{customthm}{A1}
\label{lemma_w}
For any pay scheme ${\bf w}$ satisfying Assumption \ref{ass_w},

(i) If $f$ is log-concave, then $R(e;{\bf w})\le g(x_m;{\bf v}^{\rm WTA})$;

(ii) If $f$ is log-convex, then $R(e;{\bf w})\le \frac{f(\underline{x})}{n}$.
\end{customthm}

\begin{proof}{\bf of Lemma \protect\ref{lemma_w}}

\noindent{\bf Part (i)} Let $D_1=\{{\bf x}\in\mathcal{X}^n:x_1>x_m>x_2,\ldots,x_n\}$ be the set of noise realizations such that agent 1 is the only one above the standard. Then we can write
\begin{equation}
\label{R_1}
R(e;{\bf w}) = \int_{D_1}w_1\lambda(x_1)\dd F(x_1)\ldots \dd F(x_n) + \int_{\mathcal{X}^n\setminus D_1}w_1\lambda(x_1)\dd F(x_1)\ldots \dd F(x_n),
\end{equation}
where for brevity we omit the arguments of $w_1$. In the first term in (\ref{R_1}), setting $w_1=1$ will produce an upper bound 
\begin{align*}
& - \int_{D_1}f'(x_1)\dd x_1\dd F(x_2)\ldots \dd F(x_n) = - \int_{x_m}^{\overline{x}}\dd F(x_1)\int_{\underline{x}}^{x_m}\dd F(x_2)\ldots\int_{\underline{x}}^{x_m}\dd F(x_n)\\
& = f(x_m)F(x_m)^{n-1}.
\end{align*}
Next, we show that the second term in (\ref{R_1}) is weakly increased by replacing $w_1$ with 1 for $x_1>\max\{x_2,\ldots,x_n\}$ and zero otherwise.

Suppose $x_2=\max\{x_2,\ldots,x_n\}$. Then we can split the domain of integration into $(\mathcal{X}^n\setminus D_1)\cap\{x_1>x_2\}$ and $(\mathcal{X}^n\setminus D_1)\cap\{x_1<x_2\}$. In the latter integral, we relabel variables $x_1$ and $x_2$ so that $w_1(y_1,y_2,y_3\ldots,y_n)$ turns into $w_1(y_2,y_1,y_3\ldots,y_n)$, which, by the anonymity property (a), is equal to $w_2(y_1,y_2,y_3\ldots,y_n)$. Additionally, $f'(x_1)f(x_2)$ turns into $f'(x_2)f(x_1)$, keeping everything else intact. Further, we use the budget constraint property (c) to replace $w_2$ with $1-w_1$, which weakly increases the sum. The resulting upper bound is
\begin{align*}
& \int_{\{x_1>x_2>x_m\}}w_1\lambda(x_1)\dd F(x_1)\ldots \dd F(x_n) + \int_{\{x_1>x_2>x_m\}}(1-w_1)\lambda(x_2)\dd F(x_1)\ldots \dd F(x_n) \\
& = \int_{\{x_1>x_2>x_m\}}\lambda(x_2)\dd F(x_1)\ldots \dd F(x_n) + \int_{\{x_1>x_2>x_m\}}w_1[\lambda(x_1)-\lambda(x_2)]\dd F(x_1)\ldots \dd F(x_n).
\end{align*}
Here, for brevity we omitted the remaining characterization of the integration domains and the arguments of $w_1$. Since $f$ is log-concave, $\lambda(x)$ is increasing, and hence replacing $w_1$ with 1 in the second term will increase the expression further. Using integration by parts, we can transform the result as
\begin{align*}
& \int_{\{x_1>x_2>x_m\}}\lambda(x_1)\dd F(x_1)\ldots \dd F(x_n) \\
& = - \int_{\{x_1>x_2>\max\{x_m,x_3,\ldots,x_n\}\}}f'(x_1)f(x_2)\ldots f(x_n)\dd x_1\ldots \dd x_n \\
& = - \int_{x_m}^{\overline{x}}\dd F(x_1)\int_{x_m}^{x_1}\dd F(x_2)\int_{\underline{x}}^{x_2}\dd F(x_3)\ldots \int_{\underline{x}}^{x_2}\dd F(x_n) \\
& = - \int_{x_m}^{\overline{x}}\dd F(x_1)\int_{x_m}^{x_1}\dd F(x_2) F(x_2)^{n-2}\\
& = -f(x_1)\int_{x_m}^{x_1}\dd F(x_2) F(x_2)^{n-2}\bigg|_{x_m}^{\overline{x}} + \int_{x_m}^{\overline{x}}f(x_1)^2 F(x_1)^{n-2}\dd x_1\\
& = \int_{\{x>x_m\}}F(x)^{n-2}f(x)^2\dd x.
\end{align*}

Recall that we arbitrarily chose $x_2$ to be the maximum in $\{x_2,\ldots,x_n\}$. There are $n-1$ elements in this set, all of which are equally likely to be the maximum, and all of these cases are equivalent from player 1's perspective. Thus, the expression above must be multiplied by $n-1$, which gives
\[
R(e;{\bf w}) \le f(x_m)F(x_m)^{n-1} + (n-1)\int_{\{x>x_m\}}F(x)^{n-2}f(x)^2\dd x = g(x_m;{\bf v}^{\rm WTA}),
\]
where the last equality follows directly from (\ref{g_general}) and (\ref{B_r}).

\bigskip

\noindent{\bf Part (ii)} When $f$ is log-convex, it is also convex. In conjunction with the assumption that $f(\overline{x})=0$ this implies $f$ is decreasing and $x_m=\underline{x}$. Assuming $x_2=\max\{x_2,\ldots,x_n\}$ and following the same sequence of steps as in the proof of part (i), we obtain the bound
\[
\int_{\{x_1>x_2\}}\lambda(x_2)\dd F(x_1)\ldots \dd F(x_n) + \int_{\{x_1>x_2\}}w_1[\lambda(x_1)-\lambda(x_2)]\dd F(x_1)\ldots \dd F(x_n)
\]
The log-convexity of $f$ implies that $\lambda(x)$ is decreasing, and the second term is maximized by setting $w_1=\frac{1}{n}$---the lowest value of $w_1$ that preserves the anonymity and monotonicity of the pay scheme. Then (\ref{bound_n}) gives
\[
R(e;{\bf w})\le -\frac{1}{n}\int f'(x_1)f(x_2)\ldots f(x_n)\dd x_1\ldots \dd x_n = \frac{f(\underline{x})}{n}.
\]
\end{proof}

Since the bounds on $R(e;{\bf w})$ in Lemma \ref{lemma_w} are reached by the corresponding optimal tournaments with a standard in Proposition \ref{prop_opt_tournament_with_reserve}, the result follows.
\end{proof}

\end{document}